# Interplay of Inhomogeneous Electrochemical Reactions with Mechanical Responses in Silicon-Graphite Anode and its Impacts on Degradation

Junhyuk Moon[1+], Shinya Wakita[1,3], Heechul Jung[1,3], Sungnim Cho[1,3], Jaegu Yoon[1,3], Joowook Lee[1,3], Sihyung Lee[1], Kimihiko Ito[2], Yoshimi Kubo[2], Heung Chan Lee[1+], Young-Gyoon Ryu[1,3+]

[1]Samsung Advanced Institute of Technology, 130 Samsung-ro, Yeongtong-gu, Suwon-si, Gyeonggi-do, 16678 Korea

[2]C4GR-GREEN, National Institute for Materials Science, 1-1 Namiki, Tsukuba, Ibaraki, 305-0044, Japan

[3]Present address: Samsung SDI, 130 Samsung-ro, Yeongtong-gu, Suwon-si, Gyeonggi-do, 16678 Korea

[+]Correspondence and requests for materials should be addressed to J.M. (jh.d.moon@samsung.com), H.C.L. (hchan.lee@samsung.com), and Y.-G.R. (ygryu@samsung.com).

Enhanced EV market penetration requires durability of the battery with high energy throughput. For long-term cycle stability of silicon-graphite anode capable of high energy density, the reversible redox reactions are crucial. Here, we unveil intriguing electrochemical phenomena such as crosstalk of lithium ion (Li$^+$) between silicon and graphite, Li$^+$ accumulation in silicon, and capacity depression of graphite under high pressure, which engender the irreversible redox reactions. Active material properties, i.e. the size of silicon and the hardness of graphite, silicon-graphite anode, are modified based on the unveiled results to enhance the reaction homogeneity and reduce subsequent degradation. Owing to the property change of the anode active materials, silicon-graphite anode paired with high nickel cathode allows the prismatic cell with 8.7 Ah to reach cycling performance over 750 cycles with volumetric energy density of 665 Whl$^{-1}$, which is corresponding to 800 Whl$^{-1}$ in the prismatic cell with 87 Ah. Finally, the cycling performance can be tailored by the design of electrode regulating Li$^+$ crosstalk. Our findings provide electrochemical insights into degradation mechanisms and a promising direction on the progressive improvement of materials and the design of electrodes in silicon-graphite anode.

More than 23 % of total U.S. greenhouse gas (GHG) emissions in 2016 are from dispersed automotive emissions, which is a dispersed emission sources of GHG [1, 2]. Electrification of automotive transportation is an imperative need for reducing the dispersed GHG emissions. In electrified automotive applications, the energy throughput, durability, and cost of a battery are directly associated with the successful penetration of electric vehicles to the petro-based transportation market [3]. It is widely expected that the requirements of the battery for the electric vehicles (EVs) have more than 800 Wh/L energy density at cell level for driving ranges over 500 km and cycling performance of 1300 cycles for application [3, 4, 5]. Silicon is an attractive candidate capable of achieving the required energy density because of high theoretical specific capacity, 3579 mAh/g, with moderate density for forming $Li_{3.75}Si$. However, silicon anodes suffer from rapid capacity fading owing to its large volume changes leading pulverization, loss of electrical contact, and instability of solid electrolyte interphase (SEI). Prior efforts to improve the cycling performance of silicon anodes primarily focused on resolving the chronic rapid fading issues through nano-structured silicon [6], carbon-coated/mixed silicon particle [7, 8, 9, 10, 11], and formation of stable SEI [12, 13, 14, 15]. Although their half-cell research results concluded outstanding performances over thousands of cycles, the content of the commercialized silicon anode has been limited from 2 to 3 wt % of the total anode weight so far to guarantee industrially acceptable cycle numbers [3, 16, 17, 18]. This might be due to the fact that full cell performances are not directly related to the half-cell performances in most of previous studies because of following two fundamental reasons. First, the $Li^+$ amount in a full-cell is limited by the capacity of the cathode while in a half cell it is practically unlimited. Second, the potential of an anode in a full cell is regulated by the potential difference between the cathode and the anode, which is fundamentally different from the potential of the anode versus thermodynamic reference potential such as standard hydrogen electrode (SHE) during cell operation condition. For these reasons, the locally deficient $Li^+$ could results in the cell failure in a full cell but not in a half cell, and consequently, the operation condition of anode electrode in a full-cell should be different from that in a half-cell. Therefore, identification of the origin of long-term degradation in a full-cell is decisive in improvement of the cycling performance of a complete battery with silicon anode. In an attempt to unravel the underpinning of the long-term degradation as well as to achieve commercial requirements, in this study, we designed and implemented the full-cell using the silicon-graphite anode and characterized quantitatively the states of each anode materials, silicon and graphite, under non-equilibrium operation conditions. We clearly observed the following phenomena: crosstalk of $Li^+$ between silicon and anode, $Li^+$ accumulation inside the core of silicon with cycles, and capacity depression of graphite caused by applying pressure. These phenomena closely related to their mechano-electrochemical relations have never been reported as far as we acknowledge, even in half-cell studies. Here, we focus on clarification on the relations of these phenomena and their impact on the long-term anode degradation. With an optimized design of materials and electrode that could reduce the crosstalk of $Li^+$ between silicon and graphite, we achieved the cycling performance over 750 cycles in the prismatic cell with high capacity of 8.7 Ah.

**Crosstalk of $Li^+$ between silicon and graphite:** In-operando XRD measurements provide us valuable information on the mixed state of silicon-graphite anode under non-equilibrium operation conditions. During the process of forming the stable interface between active material and electrolyte to prepare a cell for use, so called a formation process, the crystal phase of silicon transforms to amorphous phase, which is invisible to diffraction [19]. For this reason, analytical techniques should be employed to identify the $Li^+$ content in the amorphous phase of Li-Si alloy [20, 21]. We lay the analytic basis for quantification of $Li^+$ content in individual material by combining XRD data and corresponding electrochemical data. Using the data of the full cell having the graphite only anode paired with high nickel cathode, denoted as Cell A, individual $Li^+$ amount in silicon and graphite in the full cell having the silicon-graphite anode, Cell B, can be obtained as shown in Figure 1(a) (see Method for quantification of $Li^+$ content in individual material, Figure S1, and Preparation of electrode & cells in Supplementary Materials for details).

The states of charge of individual materials, denoted as individual SOCs, from XRD of full cell at 0.5C (individual SOCs, circles in Figure 1(b)) and the calculated individual SOCs from electrochemical analysis of half cells (calculated individual SOCs, solid lines in Figure 1(b)) can be compared as shown in Figure 1(b). Sudden discontinuous changes appear in the individual SOCs of silicon and graphite, marked in red triangles. The appearance of the sudden changes in the data obtained by in-operando XRD is clearer than that in the calculated individual SOCs (see Supplementary Materials for the reason of distinct signals in XRD). These correspond to discontinuities in the potentials at the phase transitions of $Li_xSi$ [22, 23, 24]. Thus, the $Li_xSi$ phase of the shell on silicon particles can be determined. $3^{th}$ position of a phase transition (red triangle at 65% SOC) of individual SOCs in Figure 1(b) corresponds to the phase transition from $Li_{2.3}Si$ to $Li_{3.25}Si$ [22, 23] that is located mostly on the surface of silicon due to the two phase transition [25]. With the assumption that the reacted volume will be maintained after $3^{rd}$ phase transition, the lithiated state of the shell of silicon particles can be estimated. Moreover, near the end of charging at SOC 87 %, the large deviation of the individual SOCs from the calculated individual SOCs is a clear signal of crosstalk between silicon and graphite. The similar electrochemical phenomena such as crosstalk between individual active materials have been reported as internal redox couple in blended cathodes [26].

The schematic of active material conditions at various SOCs are shown in Figure 1(c), which can be depicted by the analysis of in-operando XRD analysis as previously described. From 2 to 3, $Li^+$ moves from silicon to graphite corresponding to the rapid drop of silicon SOC while sudden increase of graphite SOC at the end of SOC in the Figure 1(b). Simultaneously, the pressure applied on the inner core of silicon particles might be released, and $Li^+$ in the shell could also enter into the inner core due to the released pressure. Furthermore, we can estimate the corresponding potential of individual materials, graphite and silicon, by aligning individual SOCs in to the individual potentials as shown in Figure 1(d). During charge, the potential of silicon is higher than that of graphite, and thus the crosstalk of $Li^+$ between silicon and graphite is a less probable reaction considering homogeneous reaction of each material. However, $Li^+$ in the shell on silicon particles hardly moves into the inner core of silicon because of increased pressure due to volume expansion of the lithiated shell and creates large inhomogeneity, which has already been reported as very large kinetic resistance that is a origin of the potential hysteresis of silicon [27].

Therefore, graphite particles contacting mainly with the shell of silicon might feel surface localized $Li^+$ potential of silicon particle. Considering inhomogeneous lithiation of only the shell of silicon and graphite to compare with homogeneous lithiation potential of graphite and at the cell of silicon are shown in the inset of Figure 1(d). The red line and the red points of 2 and 3 in the inlet in Figure 1(d) are derived from Gibbs free energy of formation of $Li_xSi$ [23, 28] with previously determined state of $Li_xSi$ in the shell on a silicon particle. The black line and the black points are also derived from chemical potential of formation of $Li_yC_6$ [29] with previously determined state of $Li_yC_6$, respectively (see Supplementary Materials and Figure S2 for details of thermodynamic calculation). The plots in the inlet in Figure 1(d), open circuit voltages (OCVs) of the shell of silicon particles (red line) and graphite (black line) shows inverted potential region at high capacity range that explains how internal redox reaction occurs between inhomogeneous silicon and graphite particles.

In the experiment of a half cell, we could not see internal redox couple. Lithiation in a working electrode is regulated by constant current until set potential, 0.01 V where lithium content in $Li_xSi$ reaches x=3.75 [30]. However, lithiation in an anode of a full cell is regulated by the overall potentials difference of cathode and anode. The minimum potential of an anode (~0.1 V) during constant current (CC) charge measured by 3-electrode set up was higher than that measured by a half cell experiment as shown in the upper figure of Figure 2(a). Therefore, the potential applied on an anode during constant voltage (CV) charge was lower than that in half cell.

The internal redox reaction appears even in higher C-rate of 1 C and 2 C as shown in Figure S3. At low C-rate (0.5 C, Figure 1(d)), the individual SOCs follow well with the calculated individual SOCs. When C-rate increases to 1 C and 2 C (Figure S3 (a) and (b)), the sudden changes of $Li^+$ amount due to phase $Li_yC_6$ phase transitions are smeared and the individual SOC of silicon becomes lower while that of graphite becomes higher at SOC range lower than 67 %. This might be due to poorer C-rate capability of silicon than that of graphite [31].

Li-Si alloy per se shows mechano-electrochemical coupling inside of them. Because Li-Si alloy has the convex-shaped Gibbs free energy landscape as a function of x in $Li_xSi$ with the minimum at x = 2.33 [23, 28], the spherical geometry of silicon fundamentally leads to two-phase (de)lithiation. Thus, (de)lithiation in an anode involves the interplay between electrochemical reaction and mechanical stress. $Li^+$ movements from shell to inside of a silicon particle are retarded by increased stress during charge. Because of internal redox reaction during CV charge that extracts $Li^+$ from the shell of a silicon particle, the stress applied to silicon core could be released. Simultaneously, small amount of $Li^+$ in the shell could penetrate into the pressure released core of the silicon particle. During following discharge, delithiation of the slightly lithiated inner core needs higher energetic cost due to the extended distance to the surface. For this reason, the higher the C-rate, the more $Li^+$ amount remained in silicon at the end of a cycle (Figure S4 shows the details). Consequently, internal redox reaction promotes irreversibility that facilitates $Li^+$ accumulating in the inner core of a silicon particle.

Additionally, the internal redox reaction leads to increase the time for CV charging. After CC charge is over, the high concentrated lithium at the interface between anode and electrolyte diffuses into the bulk of the anode, and the CV charge is terminated when the concentration gradient within active material is minimized [32]. There is the distinctive pathway for minimizing the concentration gradient, $Li^+$ mass transport from silicon to graphite due to the internal redox reaction. During CV charge, this pathway delays the end of CV charging time because $Li^+$ diffusivity in silicon is 2 orders lower than that in graphite [33, 34]. Consequently, the time for CV charge increases, and then delithiated cathode experiences high cut-off voltage during extended CV charging time.

**Long-term degradation:** Figure 2(a) shows that the upper plots are charge/discharge potential profiles of the silicon-graphite anode at $1^{st}$ (black line) and $250^{th}$ (red line) cycles under cycling condition of 1C rate, and the lower plots are $Li^+$ amount in each silicon (red) and graphite (black) in the silicon-graphite anode (fully delithiated) at the corresponding capacities in the upper potential profiles at $1^{st}$ (filled circle) and $250^{th}$ (open circle) cycles. The first points of the upper plots in Figure 2(a) are the OCVs before $1^{st}$ and $250^{th}$ cycles that have been changed to 0.22 V from 0.34 V, which could be caused by the remaining lithium in the anode. From the potential profile in silicon-graphite anode at slow C-rate (blue line in Figure 1(f)), $Li^+$ amount in the silicon-graphite anode could be estimated. At 0.34 V and 0.22 V, capacities of the anode are 6 % and 17.2 %, respectively. The difference is 11.2 % that agrees with 10.8 % by induced coupled plasma (ICP) measurements in Figure S5, which corresponds to the $Li^+$ amount remained in the anode after 250 cycles. As shown in Figure S5, the loss of $Li^+$ by formation of SEI mostly stems from the formation process, and its increase is smaller than an increase of the remaining $Li^+$ in anode up to 250 cycles. After discharging down to 2.8 V at 0.2 C, $Li^+$ amount in graphite is 8.1 mAh (starting point of black closed circles in the lower plot of Figure 2(a)), while $Li^+$ is almost not remained in silicon (starting point of red closed circles in Figure 2(a)). However, $Li^+$ amount remained in silicon and graphite at the end of a cycle (blue triangle in the lower plot of Figure 2(a)) are 19 and 5.1 mAh, respectively, and this remaining $Li^+$ amount is 3.7 % of charging capacity.

While $Li^+$ amount in silicon smoothly increases with one inflection point at -350 mAh at $1^{st}$ cycle during the charge from -400 to -200 mAh (red closed circles in Figure 2(a)), $Li^+$ amount in silicon at $250^{th}$ cycle (red open circles in Figure 2(a)) includes two flection points. It increases with the same slope (cyan line) of that at $1^{st}$ cycle until the the first inflection point (from x=0 to 1 in $Li_xSi$), and from the second inflection point, $Li^+$ amount in silicon begins to increase again with the same slope (green line) of that at $1^{st}$ cycle (from x=1.3 to 1.7 in $Li_xSi$). The plateau (black line on red open circles in Figure 2(a)) represent that $Li^+$ is not being inserted into silicon particles because of the existence of the accumulated $Li^+$ in the inner core of silicon particles. When the capacity reaches at -200 mAh, the anode potential drops to 0.13 V and $Li^+$ start being inserted again. All the results, the OCV differences, the remaining $Li^+$ contents from XRD at $1^{st}$ and $250^{th}$ cycles, and the $Li^+$ amount profile during charge after 250 cycles, support $Li^+$ accumulation in the inner core of silicon particles over cycles. Besides of the irreversible capacity fading due to the Li$^+$ accumulation inside of the silicon, increasing the inner volume of a silicon particle could also cause mechanical stress on the outside shell and the sequent chemical reactions of silicon-surface with electrolyte and the degradation of silicon-surface could be accelerated, as well as the accommodation site for lithium could also be reduced.

Figure 2(b) shows Li$^+$ amount in each silicon, stage 1 (LiC$_6$), and stage 2 (LiC$_{12}$) at 1$^{st}$ (filled circles) and 250$^{th}$ (open circles) cycles. Hereafter, Li$^+$ amount in stage 2 indicates the Li$^+$ amount in stage 2 and stage 2L. Li$^+$ amount in stage 1 at 1$^{st}$ cycle (grey circles in Figure 2(b)) is larger than that at 250$^{th}$ cycle (white circles in Figure 2(b)) while Li$^+$ amount in stage 2 at 1$^{st}$ cycle is smaller than that at 250$^{th}$ cycle from -200 to 200 mAh. In Figure S6, black circles and red circles are the pressures on graphite at 1$^{st}$ cycle and 250$^{th}$ cycle, and black line and red line are corresponding voltage profiles, respectively. The method for calculation of pressure on graphite is described in supplementary materials. Briefly, from the XRD peak shifts representing c-axis compression, the pressure can be calculated. Li$^+$ accumulation over cycles results in expansion of silicon, which induces that the pressure onto graphite increases during charge as shown in Figure S6. Consequently, the transition from the stage 2 to stage 1 is delayed that corresponds to the belated onset of Li increase in stage 1, as well as maximum point of staged 2 after 250$^{th}$ cycles so that the capacity of graphite loses under high pressure.

Because of previously described issues such as chemical degradation on silicon, capacity fade in graphite, and volume expansion of silicon particles, the anode materials might experience the higher current density, and the porosity in the anode electrode would change dramatically during a cycle and decrease after cycles. It was also reported that the large volume change could induce electrolyte displacement resulting in the deposition of lithium and accelerates degradation of a cell [35].

**Improvement:** Our finding brings a new twist to the design of materials and electrode. First, the optimized size of a silicon particle might increase in the homogeneity of reaction in a silicon particle, as well as their cell performance. Second, the graphite favorable for the transition from stage 2 to stage 1 under high pressure will improve the capacity retention in graphite. Lastly, reducing internal redox reaction by the design of electrode will enhance the cycle stability.

Figure 3(a) and (b) are cross-sections of the surface-treated silicon/carbon composite (SSC; see Preparation of electrodes in Supplementary Materials for more details) including silicon particles with 100 nm and 85 nm (Si 100 nm and Si 85 nm), respectively. The averaged lengths of long axis and short axis, and the equivalent radius (the radius of a sphere having equivalent volume) of silicon particles were obtained by image analysis as described in Table 1 (Image analysis in Supplementary Materials). Owing to the increased surface area, the increased oxidation of the surface of silicon entails the lower initial coulomb efficiency of the silicon-graphite anode using Si 85 nm, 88.9 % in half-cell assessment, whereas that of the anode using Si 100 nm is 90 %. However, the size reduction of silicon alleviates the Li$^+$ accumulation in inner core because of enhanced homogeneity of reaction in silicon. Figure 3 (c) and (d) show signals of X-ray absorption near edge structure (XANES) representing degradation of silicon-surface of Si 100 nm and Si 85 nm over cycles in the full cell (Preparation of electrodes & cells in Supplementary Materials). The peaks at 1843 and 1847 eV indicate lithium silicate ($Li_xSiO_y$) and silicon oxide ($SiO_x$), respectively which leads to high impedance and low specific capacity retention [36, 37, 38]. The stronger total electron yield (TEY) XANES signals at 1843 and 1847 eV strongly support that the surface degradation of silicon particles in the cell using Si 100 nm (Figure 3(c)) is more severe than that in the cell using Si 85 nm (Figure 3(d)) after 50 cycles. The partial fluorescence yield (PFY) XANES spectra show that the bulk properties of silicon particles almost do not change over cycles (Figure 3 (e) and (f)). Figure 3(g) shows the thickness change of pouch cells over time during cell cycles. The overall thickness of the cell using Si 85 nm (red line) increases less than that of the cell using Si 100 nm (black line), and Figure 3(h) shows that the cell using Si 85 nm has better cyclability than that using Si 100 nm. These results dovetail with our hypothesis. Figure 4(a) shows the pressure evolutions calculated from XRD peak (stage 1) shift on graphite in the full cell having the silicon-graphite anode using Graphite1, Cell C (black circles), and the full cell having the silicon-graphite anode using Graphite2, Cell D (red circles) Graphite1 has higher pellet density than Graphite2. (More details in Preparation of electrodes & cells in Supplementary Materials). Corresponding voltage profiles of Cell C (black line) and Cell D (red line) during a cycle are also shown in Figure 4(a). The pressure on graphite increases during charge. Figure 4(b) shows $Li^+$ amount in stage 1 and stage 2 in the cell using Graphite1 (Cell C) and that using Graphite2 (Cell D), respectively. During charge and following discharge, the maximum $Li^+$ amount in stage 2 in Cell C is larger than that in Cell D. This means that the transition from stage 2 to stage 1 in Cell C is energetically more difficult than that in Cell D under high pressure on graphite. Graphite2 with lower pellet density which corresponds to more resistive to external pressure improves the stage 2 to stage 1 transition at the higher pressure condition. Capacity fading of graphite would induce overlithiation of silicon, and this overlithiation is directly connected with internal redox reaction and $Li^+$ accumulation in silicon. The property change of active materials, e.g. size of silicon and hardness of graphite, has a strong impact on the homogeneity of electrochemical reactions in individual material. Enhancement of homogeneous reactions reduces degradation of silicon and graphite in the anode. Figure 4(c) shows the cycling performance of the prismatic cell using improved materials, Graphite2 and Si 85 nm (red circles; Preparation of cells in Supplementary materials), and the comparison of cyclability with that using Graphite1 and Si 85 nm (black circles). The prismatic cell using Graphite2 and Si 85 nm exhibits much better performance with capacity retention over 80% after 750 cycles (the coulomb efficiency of this cell can be shown in Figure S7), while the capacity that using Graphite1 and Si 85 nm steeply decreases after 200 cycles. This result clearly shows that the physical property of graphite, such as hardness, influences the cell performance. The standard capacity and volumetric energy density of this cell are 8.7 Ah and 665 Wh/L, respectively. This energy density corresponds to 800 Wh/L in the prismatic cell with capacity of 87 Ah.

So far, the anode electrode have designed on the assumption that the phase of $Li_xSi$ would be x=3.75 at full lithiation. In the conventional design of electrode, the phase of $Li_{3.75}Si$ is hard to be reached in the full cell operation condition when the internal redox reaction dominates anode reaction at the end of charge as shown in figure 1. This internal redox reaction facilitates long term degradation of a cell. We can directly modulate the internal redox reaction by the design change of the anode where the content of graphite was fixed, while the content of SSC was increased to be 111 %. The revised anode electrode consisted of SSC (18 wt %), Graphite2 (79 wt %), and binder (3 wt %). With the revised design, the cross point of individual SOCs of SSC and that of graphite becomes higher from 72.9 % to 78.8 % SOC of anode as shown in Figure 5(a) simply due to 111% increased Si content. By changing the design of electrode, the utilization of active materials in anode decreases from 91.5 % to 86.7 % (detailed calculation on the utilization of active materials in anode can be seen in Supplementary Materials). These prevent overlithiation of silicon, which is the origin of $Li^+$ crosstalk between silicon and graphite during CV charge as described previously. The increment of the accommodation site for lithium in silicon and the prevention of overlithiated silicon lead to reduction of the internal redox reaction during CV charge in the full cell. Therefore, the decrease of the internal redox reaction elicits reduction of the ratio of the time for CV charge to the total charging time as well as enhanced capacity retention as shown in Figure 5(b). In this case, Figure 5(c) and (d) show that the potentials of anode and cathode at the end of charging are almost unchanged over cycles, and so there is no concern about increased potential of cathode owing to additional silicon. Using the unveiled degradation mechanism, we can enhance the cycle performance while minimizing the adverse effects such as increased thickness, decreased initial Coulombic efficiency (ICE) of anode and decreased capacity of full cell (see cons and pros of the revised design in Supplementary Materials).

In conclusion, the internal redox reaction between silicon and graphite, and the subsequent interplay of electrochemical reaction with mechanical stress in silicon composite anode in full cell were proved by thorough mechano-electrochemical study for the first time. The reversal of the chemical potentials between surface localized $Li_xSi$ and $Li_yC_6$ drives the $Li^+$ mass transport from silicon to graphite during CV charge. This comes from that overlithiated shell of a silicon particle owing to difficulty in inserting of $Li^+$ into the inner core. Therefore, the internal redox reaction originates in the inhomogeneous reaction in silicon. Intriguingly, this internal redox reaction induces small amount of $Li^+$ to transport in the inside of a silicon particle at the end of charge because of releasing pressure in the inner core due to delithiation of the shell. Sequentially, the energy barrier and the inhomogeneity of reaction in a silicon particle lead $Li^+$ to stay in the inside of a silicon particle during discharge. As a result, $Li^+$ accumulates in the inside of silicon over cycles casing capacity fading and pressure build up. Because of this increased pressure on graphite suppresses the $Li^+$ intercalation into graphite and depresses the capacity of graphite, i.e., the inhomogeneous reaction of $Li^+$ at silicon causes internal redox reaction with graphite resulting in accumulation of $Li^+$ in silicon particles over cycles. This consequently increases the pressure applied to the graphite and decrease capacity of the graphite as well. The cycling performance is enhanced by the smaller size of silicon and the higher hardness of graphite, which increase in the homogeneity of electrochemical reactions in silicon and graphite, respectively. Moreover, the design change of anode electrode can regulate the internal redox reaction between silicon and graphite to tailor the cycling performance. Based on these mechanisms underpinning cell degradation, we realized that new design of materials and an electrode have to be required for high performance of a cell. Especially, the implementation of the high capacity prismatic cell with high volumetric energy density and decent cycle life advances the future commercialization of silicon-graphite anode. Given that the current designs of materials and electrode have enhanced the homogeneity of electrochemical reactions in silicon-graphite anode, our works will provide new perspectives on the electrochemical behavior and the design of silicon-graphite anode.

**REFERENCES AND NOTES**

[1] Greenhouse Gas Emissions, United States Environmental Protection Agency; https://www.epa.gov/ghgemissions

[2] Metz, B., et al. IPCC Special Report on Carbon Dioxide Capture and Storage. Cambridge University Press, 2005; https://www.ipcc.ch/site/assets/uploads/2018/03/srccs_wholereport-1.pdf

[3] Schmuch, Richard et al. Performance and cost of materials for lithium-based rechargeable automotive batteries *Nature Energy* **3**, 267-378 (2018)

[4] 2018 Annual Merit Review, Vehicle Technology Office (US Department of Energy, 2018); https://www.energy.gov/eere/vehicles/downloads/2018-annual-merit-review-report

[5] Andre, Dave et al. Future generations of cathode materials: an automotive industry perspective *Journal of Materials Chemistry A* **3**, 6709-6732 (2015)

[6] Liu, X. H. et al. Size-Dependent Fracture of Silicon Nanoparticles During Lithiation *ACS Nano* **6**, 1522-1531 (2012)

[7] Liu, N. et al. A pomegranate-inspired nanoscale design for large-volume-change lithium battery anodes *Nature Nanotechnology* **9**, 187-192 (2014)

[8] Son, I. H. et al. Silicon carbide-free grapheme grapwth on silicon for lithium-ion battery with high volumetric energy density *Nature Communications* **6**, 7393 (2015)

[9] Kim, S. Y. et al. Facile Synthesis of Carbon-Coated Silicon/Graphite Spherical Composites for High-Performance Lithium-Ion Batteries *ACS Applied Materials & Interfaces* **8**, 12109-12117 (2016)

[10] Ko, M. et al. Scalable synthesis of silicon-nanolayer-embedded graphite for high-energy lithium-ion batteries *Nature Energy* **1**, 16113 (2016)

[11] Wang, W. et al. Silicon and Carbon Nanocomposite Spheres with Enhanced Electrochemical performance for Full Cell Lithium Ion Batteries *Scientific Reports* **7**, 44838 (2017)

[12] Xu, C. et al. Improved Performance of the Silicon Anode for Li-Ion Batteries: Understanding the Surface Modification Mechanism of Fluoroethylene Carbonate as an Effective Electrolyte Additive *Chemistry of Materials* **27**, 2591-2599 (2015)

[13] Zhao, J. et al. Artificial Solid Electrolyte Interphase-Protected Li$_x$Si Nanoparticles: An Efficient and Stable Prelithiation Reagent for Lithium-Ion Batteries *Journal of The American Chemical Society* **137**, 8372-8375 (2015)

[14] Shen, B. H. et al. Silicon Surface Tethered Polymer as Artificial Solid Electrolyte Interface *Scientific Reports* **8**, 11549 (2018)

[15] Park, J. H. et al. Formation of Stable Solid–Electrolyte Interphase Layer on Few-Layer Graphene-Coated Silicon Nanoparticles for High-Capacity Li-Ion Battery Anodes *The Journal of Physical Chemistry C* **121**, 26155-26162 (2017)

[16] Obrovac, M. N. and Chevrier, V. L. Alloy Negative Electrodes for Li-ion Batteries *Chemical Reviews* **114**, 11444-11502 (2014)

[17] Blomgren, G. E. The Development and Future of Lithium Ion Batteries *Journal of The Electrochemical Society* **164**, A5019-A5025 (2017)

[18] Uitz, M. et al. Aging of Tesla's 18650 Lithium-Ion Cells: Correlating Solid-Electrolyte-Interphase Evolution with Fading in Capacity and Power *Journal of The Electrochemical Society* **164**, A3503-A3510 (2017)

[19] Li, J. and Dahn, J. R. An in situ X-ray diffraction study of the reaction of Li with crystalline Si *Journal of the Electrochemical Society* **154**, A156-A161 (2007)

[20] Jung, H. et al. The structural and electrochemical study on the blended anode with graphite and silicon carbon nano composite in Li ion battery *Electrochimica Acta* **245**, 791-795 (2017)

[21] Yao, K. P. C. et al. Operando quantification of (de)lithiation behavior of silicon-graphite blended electrodes for lithium-ion batteries *Advanced Energy Materials* **9**, 1803380 (2019)

[22] Wen, C. J. and Huggins, R. A. Chemical Diffusion in Intermediate Phases in the Lithium-Silicon System *Journal of Solid State Chemistry* **37**, 271-278 (1981)

[23] Liang, S.-M. et al. Thermodynamics of Li-Si and Li-Si-H phase diagrams applied to hydrogen absorption and Li-ion batteries *Intermetallics* **81**, 32-46 (2017)

[24] The potential of silicon (E) versus the amount of Li$^+$ (n) near phase transition point (E$^*$) have the form such as − u(x), where u(x) is the step function, so that asymptotic form of n versus E is $tanh^{-1}x$. The average number of Li in Li-Si alloy is $\langle n \rangle = -\partial G/\partial \mu \propto \partial(nE)/\partial E|_{E\approx E^*} \sim dtanh^{-1}E/dE|_{E\approx E^*} = 1/(1-E^2)|_{E\approx E^*}$. When phase transition of silicon occurs, the change in Li$^+$ number in Li-Si alloy can be so sudden.

[25] Wang, J. W. et al. Two-Phase Electrochemical Lithiation in Amorphous Silicon *Nano Letters* **13**, 709-715 (2013)

[26] Heubner, C. et al. Recent insights into the electrochemical behavior of blended lithium insertion cathodes: A review *Electrochimica Acta* **269**, 745-760 (2018)

[27] Sethuraman, V. A., Srinivasan, V., Newman, J. Analysis of Electrochemical Lithiation and Delithiation Kinetics in Silicon *Journal of The Electrochemical Society* **160**, A394-A403 (2013)

[28] Tipton, W. W. et al. Structures, phase stabilities, and electrical potentials of Li-Si battery anode materials *Physical Review B* **87**, 184114 (2013)

[29] Reynier, Y. et al. The entropy and enthalpy of lithium intercalation into graphite *Journal of Power Sources* **119-121**, 850-855 (2003)

[30] Hatchard, T. D. and Dahn, J. R. In Situ XRD and Electrochemical Study of the Reaction of Lithium with Amorphous Silicon *Journal of The Electrochemical Society* **151**, A838-A842 (2004)

[31] Su, X. et al. Silicon-Based Nanomaterials for Lithium-ion Batteries: A Review *Advanced Energy Materials* **4**, 1300882 (2014)

[32] Ning, G., White, R. E., Popov, B. N. A generalized cycle life model of rechargeable Li-ion batteries *Electrochimica Acta* **51**, 2012-2022 (2006)

[33] Umegaki, I. et al. Li-ion diffusion in Li intercalated graphite $C_6Li$ and $C_{12}Li$ probed by $\mu^+SR$ *Physical Chemistry Chemical Physics* **19**, 19058 (2017)

[34] Strauβ, F. et al. Lithium Tracer Diffusion in Amorphous $Li_xSi$ for Low Li Concentrations *The Journal of Physical Chemistry C* **122**, 6508-6513 (2018)

[35] Radvanyi, E. et al. Failure mechanisms of nano-silicon anodes upon cycling: an electrode porosity evolution model *Physical Chemistry Chemical Physics* **16**, 17142 (2014)

[36] Yamada, M. et al. Reaction Mechanism of "SiO"-Carbon Composite-Negative Electrode for High-Capacity Lithium-Ion Batteries *Journal of The Electrochemical Society* **159**, A1630-A1635 (2012)

[37] Schroder, K. et al. The Effect of Fluoroethylene Carbonate as an Additive on the Solid Electrolyte Interphase on Silicon Lithium-Ion Electrodes *Chemistry of Materials* **27**, 5531 (2015)

[38] Sina, M. et al. Direct Visualization of the Solid Electrolyte Interphase and Its Effects on Silicon Electrochemical Performance *Advanced Materials Interfaces* **3**, 1600438 (2016)

**Figure 1. The deconvolution of mixed state of Cell B at 0.5 C, its individual SOC, the schematic of the SOCs in individual material, silicon and graphite, and the structural change of a silicon particle, and the corresponding potentials during 1$^{st}$ charge process**

(a) The mixed state in the anode of Cell B can be deconvoluted with the individual states of silicon and graphite. The detailed voltage profile can be seen in Figure S1 (b). (b) Sudden changes of Li$^+$ amount appear in the calculated individual SOC (red triangles) and the experimental data and sudden changes correspond to the phase transformation of Li$_x$Si. Near the end of charge, experimental data gotten form full cells distinctly deviate from the calculated individual SOC obtained from a half cell. These obviously indicate the crosstalk of Li$^+$ between silicon and graphite. The discrepancy between the ends of SOCs originated from the utilization of active materials and the N/P ratio. (c) The schematic of Li$^+$ amount in silicon and graphite is able to be described by the analysis of XRD data during a cycle at 0.5 C. SOC in schematic is SOC of anode and the number in brackets is SOC of cell. The corresponding SOCs of graphite and silicon are indicated as red and blue number, respectively. The large volume expansion during insertion of Li$^+$ into silicon causes the two-phase lithiation. The formation energy of Li$_x$Si and the geometry of silicon lead to the two-phase delithiation. The mutual influence between silicon and graphite, electrochemical reactions and mechanical stress makes the structural change of silicon the several processes during a cycle. (d) The black and red circles are the corresponding potential positions of individual materials, graphite and silicon. Here, the black, red, and blue lines are the potentials of graphite, silicon, and silicon-graphite anode, respectively. During a cycle, the potential of silicon is higher than that of graphite due to two-phase lithiation. However, plots in inlet, open circuit voltages of graphite (black line) and the shell of a silicon particle (red line) show that the Li$^+$ crosstalk is possible.

**Figure 2. Li$^+$ accumulation in silicon**

The upper graph of (a) is the potential of anode, which is obtained by the 3-electrodes cell measurements. At the start of charge, OCV of 250$^{th}$ cycle is lower than that of 1$^{st}$ cycle. In the lower graph of (a), while Li$^+$ amount in silicon smoothly increases at 1$^{st}$ cycle during the charge from -400 to -200 mAh (red closed circles in Figure 2(a)), Li$^+$ amount in silicon at 250$^{th}$ cycle (red open circles in Figure 2(a)) increases with the same slope (cyan line) of that at 1$^{st}$ cycle until the begin of plateau (from x=0 to 1 in Li$_x$Si), and at the end of plateau, Li$^+$ amount in silicon begins to increase with the same slope (green line) of that at 1$^{st}$ cycle again (from x=1.3 to 1.7 in Li$_x$Si). Over cycles, Li$^+$ accumulates in silicon. At the end of discharging, Li$^+$ in silicon cannot fully move out (blue triangles). This also supports our hypothesis about the Li$^+$ accumulation. (b)Plots are Li$^+$ amount in stage 1 of graphite and silicon (the upper), juxtaposed with that in stage 2 of graphite (the lower). These show that the expansion of a Li-Si alloy due to Li$^+$ accumulation in silicon hinders the transition from stage 2 (LiC$_{12}$) to stage 1 (LiC$_6$).

**Figure 3. Size effect of silicon on degradation of silicon**

(a) and (b) are SEM micrographs of the cross-sections of SSCs. The boundaries of silicon particle are marked by red lines. The smaller the size of a silicon particle becomes, the smaller the accumulated Li$^+$ in the inside of a silicon particle. In XANES signals, the increase of peaks at 1843 and 1847 eV indicates the evolution of lithium silicate (Li$_x$SiO$_y$) and silicon oxide (SiO$_x$), respectively. The evolution of lithium silicate and silicon oxide leads to the high impedance and low specific capacity retention. The TEY XANES strongly support that (c) the surface degradation of silicon particles in the cell using Si 100 nm becomes severer than (d) that in the cell using Si 85 nm after cycles. (e), (f) The PFY XANES show that the bulk properties of silicon particles almost do not change after cycles. The accumulated Li$^+$ in the inside of silicon directly influences on the surface degradation. Consequently, the cell expansion is reduced after cycling. (g) The thickness of the cell using Si 85 nm increases less than that of the cell using Si 100 nm. Furthermore, expanding of silicon by Li$^+$ accumulation in its inner core leads to mechanical stress on the outer shell of silicon particles. This accelerates not only the chemical degradation of silicon-surface, but also the capacity fade of graphite due to the volume expansion of silicon. (h) the cell with Si 85 nm has better performance than that with Si 100 nm.

**Figure 4. Capacity depression of graphite under high pressure due to the volume expansion of Li$_x$Si**

For comparison of Cell C with Cell D, the start time of discharge is set to 0 minute. (a) The pressure on graphite evolves by the expansion of Li$_x$Si and the sequent Li$^+$ intercalation into graphite. (b) During charge, the maximum Li$^+$ amount in stage2 in Cell C is larger than that in Cell D. This indicates that the transition from stage2 (LiC$_{12}$) to stage1 (LiC$_6$) in Cell C is energetically more difficult than that in Cell D. (c) These prismatic cells with capacity of 8.7 Ah have the volumetric energy density of 665 Wh/L, which corresponds to 800 Wh/L in EV cell with capacity of 87 Ah.

**Figure 5. Change of the design of electrode**

The graph (a) is the comparison of the calculated individual SOCs between prior and revised designs. In the revised design, the cross point of individual SOCs of SSC and graphite is higher SOC than that in the prior design, and the utilization of active materials in anode decreases, i.e. the revised design can reduce the internal redox reaction. The performance of the cell having newly designed anode electrode is compared with that of the prior cell. The revised cell shows the better performance such as (b) capacity retention and ratio of the time for CV charge to total charging time (CV ratio). The graph shown in (c) is the individual potentials of a cathode and an anode at the end of charge and at the end of charge over cycles. The potential profiles shown in (d) are the potential of full cell and the individual electrode potentials at 50$^{th}$ cycle.

|  | **Avg. length of long axis** | **Avg. length of short axis** | **Equivalent radius** |
|---|---|---|---|
| **Si 100 nm** | 99.91 nm | 37.76 nm | 27.61 nm |
| **Si 85 nm** | 85.6 nm | 33.25 nm | 24.21 nm |

**Table 1. the averaged lengths of long axes, short axes and the equivalent radii obtained from SEM micrographs.**

# SUPPLEMENTARY MATERIALS

**Preparation of Electrodes:** SSC was prepared as follows: silicon particles were prepared by ball-milling process with micron-sized silicon powder followed by chemical vapor deposition to coat carbon on the silicon particle using a custom-made rotary kiln. An additional surface treatment was carried out with coal tar pitch as the carbon source by mixing in a powder mixer and heating at 900 °C for 3h to complete the carbonization reaction. This coated structure allows highly conductive percolation as well as retardation of unwanted reaction between silicon and electrolyte. The size of the synthesized SSC particles was in the range of 5 to 15 μm observed by a Scanning Electron Microscope (SEM, Helios NanoLab 450HP, FEI). The specific capacity of SSC was 1480 mAh/g, which was measured using lithium half-cells with the SSC as the working electrode at a delithiation rate of 0.2 C. Here, the areal capacity of electrode was 1.5 mAh/cm$^2$. The operation scheme for testing anode material via half cells is CC-CV lithiation and CC delithiation modes between 1.5 and 0.01 V with a 0.05 C cutoff at 25 °C. This measured capacity of SSC was statistically the same as the theoretical capacity of the silicon embedded in the SSC. Hence, capacity portions of the carbon and oxides in SSC could be ignored. The content of silicon in SSC was crosschecked by three different analyses, inductively coupled plasma-atomic emission spectroscopy (ICP-AES, ICPS-8100, Shimadzu), thermal gravimetric analysis (TGA, TGA/DSC2, METTLER TOLEDO), and Oxygen Nitrogen Analyzer (EMGA-920, HORIBA).

The anode electrode was fabricated by roll-to-roll coating of slurry dissolved in deionized water on cupper foil having 8 μm thickness. The slurries for the areal capacity of 4.4 and 6 mAh/cm$^2$ consisted of SSC (14.9 wt%), graphite (82.1 wt%), and binder (3 wt%), and SSC (16.5 wt%), graphite (80.5 wt%), and binder (3 wt%), respectively. Here, we used two variation of graphite, Graphite1 and Graphite2. The pellet densities of Graphite1 and Graphite2 were 1.62 g/ml and 1.47 g/ml, respectively. The pellet was fabricated by pressing graphite powder under pressure of 15.4 MPa for 30 seconds. The electrode was dried under vacuum at 130 °C for 3 h and roll-pressed at room temperature in a custom-made line. The specific capacity of the SSC and graphite electrode as the working electrode was 520 mAh/g.

Li$_{1.0}$Ni$_{0.88}$Co$_{0.08}$Mn$_{0.04}$O$_2$ Powder (NCM) was prepared by mixing co-precipitated Ni$_{0.88}$Co$_{0.08}$Mn$_{0.04}$(OH)$_2$ powder with LiOH·H$_2$O and calcinating the mixture at 750°C for 40 h in O$_2$ in a custom-made Roller Hearth Kiln (RHK). The washing process involved stirring the prepared NCM powder in deionized water with a mechanical stirrer for 10 min, followed by filtration within 5 min. The weight ratio between the NCM and the washing water was 1:3 over 10 min. The remaining water was evaporated in an air convection oven at 720 °C overnight followed heat treatment at 720 °C for 5 h in flowing O$_2$ gas using RHK. The specific capacity of NCM measured by half cell discharge was 214.8 mAh/g. The operation scheme of the half cell with the cathode material was CC-CV charging and CC discharging modes at 0.2 C between 4.3 and 2.8 V with a 0.05 C cutoff at 25 °C.

The cathode electrode was fabricated by roll-to-roll coating of a slurry consisting of NCM (96 wt%), Carbon black (2 wt%, Cabot Co.), and polyvinylidene difluoride (PVDF, 2 wt%, Solef) in N-methyl-2-pyrrolidone (NMP, Sigma-Aldrich) on aluminum foil with 10μm thickness. The electrode was roll-pressed and dried under vacuum at 120°C in a custom-made line.

**Preparation of Cells:** Pouch cells of four different electrodes chemistries were tested for in-operando XRD experiments. Hereafter, they will be called as Cell A, B, C and D. Cell A consists of NCM cathode and Graphite1 anode with 4.4 mAh/cm$^2$ and N/P ratio of 1.08. Cell B consists of NCM cathode and SSC-graphite1 anode, with the areal capacity of 4.4 mAh/cm$^2$. The N/P ratio, areal capacity ratio between anode and cathode was 1.03, Cell C NCM cathode and SSC-Graphite1 anode with 6 mAh/cm$^2$ and 1.03 N/P ratio, and Cell D NCM cathode and SSC-Graphite2 anode with 6 mAh/cm$^2$ and 1.03 N/P ratio. The pouch cell composed of two double-side-coated cathodes of 33.4 x 87 mm$^2$ size, one double-side-coated anode, and two single-side-coated anodes of 34.4 x 88.5 mm$^2$ size separated by a ceramic-coated polyethylene porous film (CCS, Toray). The capacities of Cell A and Cell B were 430 mAh and 480 mAh, respectively, and that of Cell C and Cell D were 630 mAh. All the cells were filled with 1.15 M lithium hexafluorophosphate (LiPF$_6$, Panax E-Tec) dissolved in fluoroethylene carbonate (FEC, Panax E-Tec), ethylene carbonate (EC, Panax E-Tec), ethylmethyl carbonate (EMC, Panax E-Tec), and dimethyl carbonate (DMC, Panax E-Tec) mixed solvent in 7:7:46:40 by volume, respectively.

For the comparison of size effects of silicon particles on volume expansion and cyclability, the anodes having the 85 nm and 100 nm silicon particles embedded in SSCs were tested. The anode electrodes consisted of SSC (14.6 wt%), graphite (82.4 wt%), and binder (3 wt%). For the in-situ thickness measurements, pouch cells were fabricated using the double-side-coated cathode of 32 x 40 mm$^2$ size, and the single-side-coated anode of 35 x 85 mm$^2$ size which was folded in half length-wise placing the prepared cathode inside with the CCS in-between the cathode and the anode. The cathode powders consisted of Li$_{1.0}$Ni$_{0.6}$Co$_{0.2}$Mn$_{0.2}$O$_2$ (80 wt%) and Li$_{1.0}$Ni$_{0.8}$Co$_{0.1}$Al$_{0.1}$O$_2$ (20 wt%). These cells were filled with 1.15M LiPF$_6$ dissolved in EC, EMC, and DMC in 2:4:4 by volume. The areal capacity of these cells was 3.4 mAh/cm$^2$ and the N/P ratio was 1.03. Cyclability performance of the anode materials were carried out using so called mini-18650 cells, which are cylindrical-type 18650 cell with a hollowed-out cylindrical polypropylene (PP) insert to reduce the effective internal cell volume and total capacity of the cell. The mini-18650 cell composed of one double-side-coated cathode of 54 x 200 mm$^2$ size and one double-side-coated anode of 57 x 240 mm$^2$ separated by the CCS. The cathode powders consisted of Li$_{1.0}$Ni$_{0.6}$Co$_{0.2}$Mn$_{0.2}$O$_2$ (80 wt%) and Li$_{1.0}$Ni$_{0.8}$Co$_{0.1}$Al$_{0.1}$O$_2$ (20 wt%). These cells were filled with 1.15 M LiPF$_6$ dissolved in FEC, EC, EMC, and DMC in 7:7:46:40 by volume. The areal capacity and N/P ratio of these cells were 3.4 mAh/cm$^2$ and 1.03, respectively. The prismatic cell composed of 23 double-side-coated cathodes and 24 double-side-coated anodes separated by CCS in a can of 10 x 80 x 60 mm$^3$ size. The electrodes in this prismatic cell were almost the same with Cell D, but the silicon particle embedded in SSC of anode electrodes was Si 85 nm. The standard capacity, gravimetric energy density, and volumetric energy density of this cell are 8.7 Ah, 318 Wh/kg, and 665 Wh/L, respectively. This energy density corresponds to 800 Wh/L in the prismatic cell of 87 Ah capacity. All cell fabrication, including the electrode coating and assembly of pouch cells, was conducted in a dry room with dew point below -40 $^{\circ}$C.

**Electrochemical Evaluation:** All pouch cells were placed between two aluminum plates under initial pressure of 35 Pa, and were connected with the battery cycle tester (TOSCAT, Toyo). After formation process and two standard cycles, the cells were cycled at 25 $^{\circ}$C, with a 10 minutes rest period between cycles. For each cycle, the cells with the areal capacity of 4.4 mAh/cm$^2$ (Cell A and Cell B) and 6 mAh/cm$^2$ (Cell C and Cell D) were charged up to 4.3 and 4.25 V, respectively at the cycle rates of 1 C and 0.5 C and held in the constant voltage (CV) charging mode, which was stopped after the charging current was less than 0.05 C. After 10 minute resting period, Cell A and B, and Cell C and D were discharged down to 2.8 V at the C-rate of 1 C and 0.5 C, respectively. The large cell performance was characterized using prismatic cell. The test condition of this cell was identical to that of Cell D.

A three-electrode pouch cell was prepared to characterize the individual electrochemical behavior of the cathode and the anode of a full cell. Lithium foil was placed in the middle of the bottom in order to minimize the over-potential [1]. During cycling, the individual electrode potentials were recorded versus lithium reference electrode.

For the mini-18650 cells, the formation and standard schemes of the cells were 0.1 C and 0.2 C CC-CV charging and CC discharging modes, respectively between 4.2 and 2.8V with a 0.05 C cutoff at 25 $^{\circ}$C. The cells were cycled using a 1 C CC-CV charging and CC discharging modes between 4.2 and 2.8 V at 25 $^{\circ}$C.

During cycling tests, DC-IR was measured every 100 cycles. For DC internal resistance (DC-IR) measurement, all cells were charged up to SOC 50 at the C-rate of 0.2 C. After the measurement, all cells were charged up to the upper cutoff voltage at the C-rate of 0.2 C, and the cycling tests were continuously carried out

**Thickness Measurements:** In-situ thickness measurements of the single-stack pouch cells during cycling were conducted using a load cell tester. After imposing of a 2kg SUS block on the top of the cells, the cells were connected with the battery cycle tester. The formation and standard schemes of the cells were 0.1 C and 0.2 C CC-CV charging and discharging modes, respectively between 4.2 and 2.8 V with a 0.05 C cutoff at 25 $^{\circ}$C. The cells were cycled using a 0.5 C CC-CV charging and discharging modes between 4.2 and 2.8 V at 25 $^{\circ}$C. The thickness of the cell was monitored with a gap sensor of non-contact displacement measuring system (Linear Gauge System, Mitutoyo) during cycling.

**In-Operando XRD:** The high x-ray intensity for the operando measurements was achieved with the X-ray diffractometer installed at BL15XU in SPring-8 (Hyogo, Japan). The pouch cell was fully discharged down to 2.8 V with 0.2 C rate, and placed between two aluminum plates having beryllium window in the same way described in the electrochemical evaluation section. The photon energy used for transmission through a pouch cell was 18.987 KeV, of which wavelength was 0.65297 Å. All diffraction profiles were collected using 4 detectors (1D multichannel) in the range $0 < 2\theta < 50$ °. Sequential XRD profiles were obtained by exposure during 10 seconds every 1 minute for the operation of 1 C and 2 C rates, and every 2 minutes for the operation of 0.5 C rate. The profiles containing the diffraction peaks of graphite and current collector, Cu in the range of $1.45 < q < 2.0$ Å$^{-1}$ were extracted, and were fitted using OriginPro® to deconvolute and determine the phases of the graphite and their peak intensities. The previous crystallographic data of graphite [2, 3, 4, 5, 6] were referred for the fitting. The electrochemical data were measured simultaneously.

**XAFS:** Ex-situ Si K-edge XAFS was measured at BL-10 in Ritsumeikan SR center. After fully discharging the cell, the cells were disassembled in an Ar-filled glovebox. The anodes were rinsed with DMC for 5 minutes, set on carbon-taped specimen holders, loaded in to an airtight vessel, and then transferred to the BL-10 chamber without exposure to ambient air. The vessel was immediately evacuated, and the specimens were loaded into the measurement chamber with a vacuum level of $5 \times 10^{-8}$ Pa. The photon beam energy was from 1000 to 2500 eV with a resolution of 0.5 eV or less. The Si K-egde x-ray absorption near edge structure (XANES) of an anode was measured using the total electron yield (TEY, probing depth ~nm) and partial fluorescence yield (PFY, probing depth ~10-10$^2$ nm) [7, 8].

**Calculation of Lithium-ion Contents in SSC and Graphite:** Lithium half-cells with SSC, graphite, and composite of SSC and graphite as a working electrode were composed, respectively. The areal capacity of electrode was 1.5 mAh/cm$^2$. Their charge/discharge potentials were measured at 0.05 C, which was presumed to be slow enough to measure pseudo equilibrium potential. Considering the mass ratio of SSC and graphite, the Li$^+$ contents in each component in the composite electrode at a certain potential were calculated by using the respectively measured SOC vs. potentials profile of the SSC and graphite. The calculated SOC vs. potential profile for composite electrode was well matched with the measured one. Through this process, the individual SOCs of each material could be separated and simulated in the various compositions.

**Image Analysis:** The cross sections of anodes were prepared using the ion beam cross section polisher combined with nitrogen cooling system and air isolation system (IB-19520CCP, JEOL), in order to prevent ion-beam damage and exposure to ambient air. These specimens were then transferred to SEM (Helios NanoLab 450HP, FEI) using the air isolation transfer system. The silicon nano-particles embedded into carbon in SSC were imaged in the Secondary Electron (SE) mode at an accelerating voltage of 2 kV, and magnification at x 250000.

Sets of features corresponding to the silicon boundaries needed to be extracted from the gray-scaled SEM micrographs. These grays-scaled images were converted to binary images, and the coordinates associated with the boundaries could be estimated. In order to preserve silicon boundaries and edges having large changes in intensity while selectively removing uncertain lines with small intensity fluctuations, L0 gradient minimization [9] (with weighting factor 0.009) was used, which is particularly effective for highlighting silicon boundaries. And morphological processing [10] (dilation / erosion) was used to preserve original object shapes. Otsu's method [11] was used to segment silicon (white, "1") and others (black, "0"). Note that, when silicon was overlapped in the image, silicon particles were partially divided manually. Finally MATLAB® 2018a built-in function "regionprops" was conducted to calculate the average particle area and diameter. All algorithms were implemented using MATLAB® 2018a (The MathWorks, Inc.).

**Method for quantification of Li$^+$ content in individual material:** In this work, the coulomb efficiencies of full-cells reach as high as over 99.8 %, so we ignore lithium loss due to the SEI formation and the deterioration of electrodes for a single cycle. Thus, the Li$^+$ amount is considered to be in only graphite and silicon. We are able to describe the following linear equation based on the assumption that XRD intensity of each phase is linearly proportional to the Li$^+$ content of each phase.

$$\begin{pmatrix} A_i & \cdots & E_i \\ \vdots & \ddots & \vdots \\ A_f & \cdots & E_f \end{pmatrix} \begin{pmatrix} \alpha \\ \vdots \\ \varepsilon \end{pmatrix} + \begin{pmatrix} \varphi_i \\ \vdots \\ \varphi_f \end{pmatrix} = \begin{pmatrix} Capacity_i \\ \vdots \\ Capacity_f \end{pmatrix} \qquad (S1)$$

where, $A_k$, $B_k$, $C_k$, $D_k$, and $E_k$ are intensities of peaks for stage1, stage2L, stage2, stage3L, and stage4L, respectively, which are extracted from the fit results of XRD, and $\alpha$, $\beta$, $\gamma$, $\delta$, and $\varepsilon$ are their coefficients converting the intensity of each phase into its capacity, respectively. $\varphi_k$ is the Li$^+$ content in silicon, and $Capacity_k$ is the capacity of a cell in the k$^{th}$ sequential data. Herein, the stages are named by the ordering of lithium ions in every n$^{th}$ interslab of graphite, and the letter L indicates liquid-like, without any in-plane order [12].

According to Equation (S1), the Li$^+$ amount in silicon can be reconstructed. Firstly, the coefficients, $\alpha$, $\beta$, $\gamma$, $\delta$, and $\varepsilon$, were extracted from the in-oprando XRD results of the full cell having the graphite only anode paired with high nickel cathode, denoted as Cell A (see Preparation of electrodes & cells in Supplementary Materials for more details), by exploiting the pseudo-inversion method [13] (Figure S1(a)). Secondly, XRD profiles were normalized using the intensity of Cu(111) peak. Then, we could calculate Li$^+$ amount in only graphite in the full cell having the silicon-graphite anode, Cell B (Figure S1(b); Preparation of electrodes & cells in Supplementary Materials). Finally, the Li$^+$ amount in silicon in Cell B could be obtained (Figure S1(c)).

**Why do the individual SOCs obtained from XRD results show clearly the sudden changes?** The peaks in dQ/dV of Li-Si alloy are typically much broader than that of graphite, which indicates that various phases of Li-Si alloy exist in a silicon particle at the same time. Owing to the increasing pressure due to lithiation, the retardation of Li$^+$ insertion into the inside of silicon during charge could promote the gradient of phases in Li-Si alloy. Because we used a small sized silicon particle, of which the length of short axis is about 30 nm, the gradient of the phases in Li-Si alloy could be small during a cycle. According to the argument in Note (24), the change in the amount of Li$^+$ is more sudden than the change in the potential near phase transition. Finally, the numbers of Li$^+$ are directly associated with the diffraction peaks measured by XRD. For these reasons, the sudden changes in the individual SOCs obtained from XRD results are clearly shown in our work.

**Why internal redox reaction occurs?** The phase of Li$_x$C$_6$ could be determined by Li$^+$ amount in graphite and the phase of the shell of silicon particles could be also determined using the position of a phase transition from the XRD results. It is hard for Li$^+$ in the shell on silicon particles to move into the inner core of silicon because of increased pressure due to volume expansion of the lithiated shell. We can assume that graphite interacts only with the shell of silicon particles. The difference of chemical potentials can be described at constant temperature and pressure by the following equation:

$$\Delta\mu = \left(\mu_{Li_yC_6} - \mu_{Li}\right) - \left(\mu_{Li_xSi} - \mu_{Li}\right) = \left(\left.\frac{\partial G_{Li_yC_6}}{\partial y}\right|_{T,P} - G_{Li}\right) - \left(\left.\frac{\partial G_{Li_xSi}}{\partial x}\right|_{T,P} - G_{Li}\right) \tag{S2}$$

where $\mu_{Li_xSi}$, $\mu_{Li_yC_6}$, $\mu_{Li}$, $G_{Li_xSi}$, $G_{Li_yC_6}$, and $G_{Li}$ are the chemical potentials of Li$_x$Si, Li$_y$C$_6$, and metallic Li and the corresponding Gibbs free energies of formation of Li$_x$Si, Li$_y$C$_6$, and metallic Li [14], respectively [15]. The chemical potential is the change in the corresponding Gibbs free energy that occurs as Li$^+$ amount changes. In the case of charging Cell A at 0.5 C, 3$^{rd}$ position of a phase transition (red triangle at 65 % SOC) of individual SOCs in Figure 1(d) corresponds to the phase transition from Li$_{2.3}$Si to Li$_{3.25}$Si that is located mostly on the surface of silicon due to the two phase lithiation [16]. The electrochemically active volume, presumably surface, to total volume of the silicon particle can be calculated as 70.7 % from the difference between the surface phase, Li$_{3.25}$Si, and the measured total Li$^+$ content in silicon at 3$^{rd}$ position of a phase transition. With the assumption that the reacted volume will be maintained after 3$^{rd}$ phase transition, the lithiated state of the shell of silicon particles can be estimated and, subsequently, the chemical potentials of $\mu_{Li_xSi}$ and $\mu_{Li_yC_6}$ can also be calculated using the previously reported thermodynamic values [17, 18, 19]. Here, the change in pressure is ignored and the pressure and temperature of this system are set to 1 atm and 297 K. Figure S2 summarizes all of these processes for calculating the thermodynamic driving force during CV charge and rest. In Figure S2, the most upper figure is the driving force for the internal redox couple. Li$^+$ can move from silicon to graphite during CV charge. Our situation is different from the conventional solid-solution model for homogenous states, but is well described by Bazant's report for nonequilibrium and inhomogeneous state [20].

**Measurement of lithium in anode:** After discharging the cell, the cells were disassembled in an Ar-filled glovebox. For measuring the electrochemically inactive lithium amount, the half-cell was assembled using the disassembled anode, fully delithiated up to a cutoff voltage, 1.5 V at the C-rate of 0.05 C, and dissembled the half-cell in as Ar-filled glovebox. The anodes were rinsed with DMC for 5 minutes, and dried. After chemical treatments [21], the lithium amount in anode was determined using ICP-AES. The lithium difference in the discharged anode and the reassembled-and-delithiated anode could be the remaining $Li^+$ in anode. After 1st cycle and 250th cycles of Cell A, we measured total lithium (black columns), inactive lithium (red columns), and the remaining $Li^+$ (blue columns) in anode as shown in Figure S5. Here, the inactive lithium can be considered as the consumed lithium for forming SEI. Using loading level (g/cm$^2$) of anode, molar masses of carbon, silicon, and lithium, and the atomic composition in anode, the SOCs of anode after 1st and 250th cycles can be calculated as 3.5 and 14.3 %, respectively.

**Calculation of pressure on graphite:** When the lithiation of stage 1 is relatively small, the fitting error of peak's position is larger due to small signal. From XRD peak shift, the pressure applied on graphite can be calculated using the following equation.

$$\Delta P \approx \frac{1}{\kappa_c} \frac{c_0 - c}{c_0} \tag{S3}$$

where, the compressibility of the c axis of $LiC_6$ is reported to be $\kappa_c = 1/c_0(\partial c/\partial P)_{P_0} = 1.344 \times 10^{-2}$ GPa$^{-1}$, c is the c-axis constant, and $c_0$ is the c-axis lattice constant at the reference pressure $P_0$ [22].

**Utilization of active materials in anode:** The discharge capacity of the full cell, $Q_c$, is following as

$$Q_c = Q_0^+ - Q_I^- = Q_0^+ - \frac{n}{p} Q_0^+ (1 - \varphi^-) \tag{S4}$$

where, $Q_0^+$, $Q_I^-$, n/p, and $\varphi^-$ are the initial cathode charge capacity, irreversible capacity of anode, negative to positive capacity ratio, and initial Coulombic efficiency (ICE) of anode, respectively. Given by the utilization of cathode, α, by the lower cutoff voltage, the utilization of anode, β, can be written as

$$\beta = \frac{\alpha Q_c}{Q_R^-} = \frac{\alpha Q_c}{\frac{n}{p} Q_0^+ \varphi^-}. \tag{S5}$$

where, $Q_R^-$ is the reversible capacity of anode [23].

By changing the design of electrode, N/P ratio increases from 1.03 to 1.08, and the ICE of anode decreases from 88.9 % to 88.7 %. Then, the discharge capacity of the full cell decreases from 88.57 % to 87.8 % of the initial cathode charge capacity after the formation process. Given by the lower cutoff potential of cathode, 3.35 V, the utilization of cathode is 94.6 %. Therefore, the utilization of anode decreases from 91.5 % to 86.7 %.

**Cons and Pros of the revised design:** The thicknesses of anode in the prior design and revised design are 73.3 μm and 76.8 μm, respectively. By changing the design of electrode, its thickness increases by 5 %, but the change of thickness at the full volume expansion decreases by 2.5 % (from 123 to 120.5 % vs. initial thickness). Simultaneously, the discharge capacity of 1st cycle after the formation process decreases by 0.9 % (from 606 to 600.5 mAh), but the capacity retention ratio increases from 95.4 to 97.1 % after 100 cycles.

**REFERENCES**

[1] Mitsuda, K., Hara, S., Takemura, D. Polarization Study and AC-Impedance Measurement in the Horizontal Plane of a Lithium-ion Battery Using Eight Reference Electrodes *Electrochemistry* **84**, 861-871 (2016)

[2] Ohzuku, T., Iwakoshi, Y., Sawai, K. Formation of Lithium-Graphite Intercalation Compounds in Nonaqueous Electrolytes and Their Application as a Negative Electrode for a Lithium Ion (Shuttlecock) Cell *Journal of The Electrochemical Society* **140**, 2490-2498 (1993)

[3] Billaud, D., Henry, F. X. Structural studies of the stage III lithium graphite intercalation compound *Solid State Communications* **124**, 299-304 (2002)

[4] Billaud, D., Henry, F. X., Lelaurain, M., Willmann, P. Revisited Structures of Dense and Dilute Stage II Lithium-Graphite Intercalation Compounds *Journal of Physics and Chemistry of Solids* **57**, 775-781 (1996)

[5] Woo, K. C. et al. Effect of In-Plane Density on the Structural and Elastic Properties of Graphite Intercalation Compounds *Physical Review Letters* **50**, 182-185 (1983)

[6] Guerard, D., Herold, A. Intercalation of lithium into graphite and other carbons *Carbon* **13**, 337-345 (1975)

[7] Jiang, D. T. et al. Observations on the surface and bulk luminescence of porous silicon *Journal of Applied Physics* **74**, 6335 (1993)

[8] Sham, T. K. et al. Electronic structure and optical properties of silicon nanowires: A study using x-ray excited optical luminescence and x-ray emission spectroscopy *Physical Review B* **70**, 045313 (2004)

[9] Xu, L., Lu, C., Xu, Y., Jia, K. Image smoothing via L0 gradient minimization *ACM Transactions on Graphics* **30**, 174-184 (2011)

[10] Serra, J. Image Analysis and Mathematical Morphology, vol. 1. London: *Academic Press*, 1982

[11] Otsu, N. A threshold selection method from gray-level histograms *IEEE Transactions on Systems, Man, and Cybernetic* **SMC-9**, 62–66 (1979)

[12] Heβ, M., Novák, P. Shrinking annuli mechanism and stage-dependent rate capability of thin-layer graphite electrodes for lithium-ion batteries *Electrochimica Acta* **106**, 149-158 (2013)

[13] Penrose, R. On best approximate solution of linear matrix equations *Proceedings of the Cambridge Philosophical Society* **52**, 17-19 (1956)

[14] Barin, I., Platzki, G. Thermochemical data of pure substances, 3$^{rd}$ ed. New York: VCH: Weinheim, 1995

[15] Aydinol, M. K. et al. Ab initio study of lithium intercalation in metal oxides and metal dichalcogenides *Physical Review B* **56**, 1354-1365 (1997)

[16] Wen, C. J. and Huggins, R. A. Chemical Diffusion in Intermediate Phases in the Lithium-Silicon System *Journal of Solid State Chemistry* **37**, 271-278 (1981)

[17] Tipton, W. W. et al. Structures, phase stabilities, and electrical potentials of Li-Si battery anode materials *Physical Review B* **87**, 184114 (2013)

[18] Liang, S.-M. et al. Thermodynamics of Li-Si and Li-Si-H phase diagrams applied to hydrogen absorption and Li-ion batteries *Intermetallics* **81**, 32-46 (2017)

[19] Reynier, Y. et al. The entropy and enthalpy of lithium intercalation into graphite *Journal of Power Sources* **119-121**, 850-855 (2003)

[20] Bazant, M. Z. Theory of Chemical Kinetics and Charge Transfer based on Nonequilibrium Thermodynamics *Accounts of Chemical Research* **46**, 1144-1160 (2013)

[21] Mao, Z. et al. Calendar Aging and Gas Generation in Commercial Graphite/NCM-LMO Lithium-Ion Pouch Cell *Journal of The Electrochemical Society* **164**, A3469-A3483 (2017)

[22] Kganyago, K.R., Ngoepe, P. E. Structural and electronic properties of lithium intercalated graphite $LiC_6$ *Physical Review B* **68**, 205111 (2003)

[23] Obrovac, M. N. and Chevrier, V. L. Alloy Negative Electrodes for Li-ion Batteries *Chemical Reviews* **114**, 11444-11502 (2014)

**Figure S1. The deconvolution of mixed state of Cell B at 0.5 C**

(a, b) X-ray diffraction profiles of anodes and the corresponding electrochemical data during battery operation. The amplitude profiles of the stages of graphite in an anode after refinement process are indicated on the right of each XRD corresponding with voltage and current profile. The coefficients for converting the amplitude of each stage to its capacity are obtained from data of Cell A by exploiting the pseudo-inversion method. The amplitude profiles of the stages of graphite in Cell B are converted to their profiles of $Li^+$ amount using the calculated coefficients. Sequentially, the profile of $Li^+$ amount in silicon can be obtained. (c)The mixed state in the anode of Cell B can be deconvoluted with the individual states of silicon and graphite.

**Figure S2. The thermodynamic driving force matching with XRD results during CV charge and rest time of Cell B at 0.5 C**

The start time of CV charging mode is set to 0 minute. After 17 minutes, the cell is in rest. The phase of $Li_xC_6$ could be determined and the phase of the shell of silicon particles could be estimate using the position of sudden change in $Li^+$ amount. The difference of chemical potentials between $Li_xC_6$ and $Li_xSi$ becomes increasing because of the internal redox couple between silicon and graphite

**Figure S3. The SOC in individual material during charge at 1 C and 2 C**

(a) and (b) are comparisons between the individual SOC and the calculated individual SOC in the anode of Cell B during charge at 1 C and 2 C, respectively. The lines were drawn based on the experiments of half cells at low rate, 0.05 C. The red and black circles were obtained by experiments of full cells.

**Figure S4.** $Li^+$ amount in silicon and graphite during 1st charge and discharge processes at various C-rates. The positions indicated by inversed triangles are marked as the $Li^+$ amounts remained in silicon at the end of a cycle, i.e., the dark blue inversed triangle 57 mAh at 2C, the blue 19 mAh at 1C, and the light blue 0 mAh at 0.5C.

**Figure S5.** Total lithium amount, inactive lithium amount, and the remaining $Li^+$ in anode of Cell B after 1st cycle and 250th cycle.

**Figure S6. Evolution of pressure on graphite by the volume expansion of $Li_xSi$ and corresponding voltage profiles**

$Li^+$ accumulation over cycles results in expansion of silicon, which induces that the pressure onto graphite increases.

**Figure S7.** The Coulombic efficiencies of the prismatic cell with capacity of 8.7 Ah

(Inset) Magnified view of the Coulombic efficiencies in the first 50 cycles.

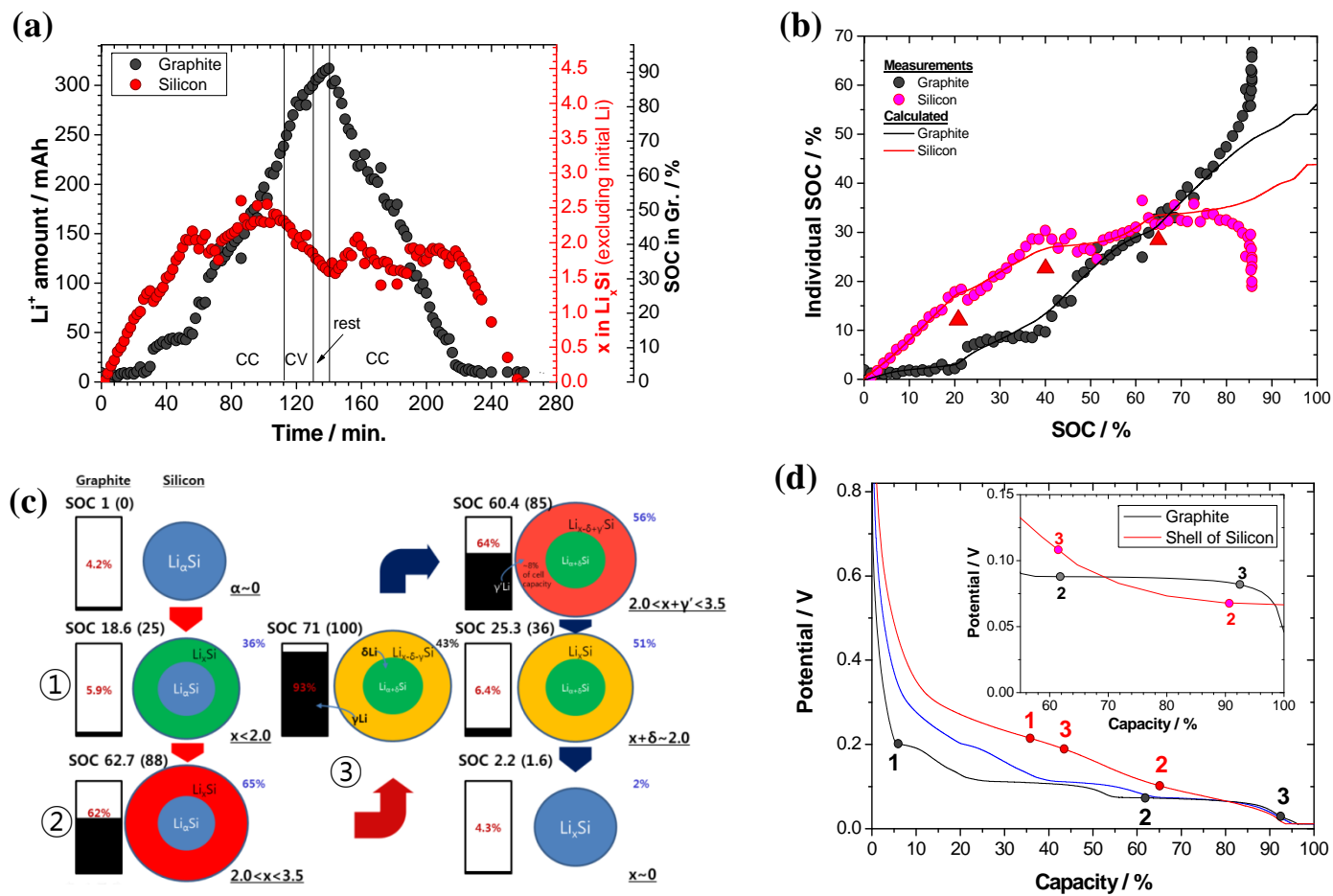

**Figure 1**

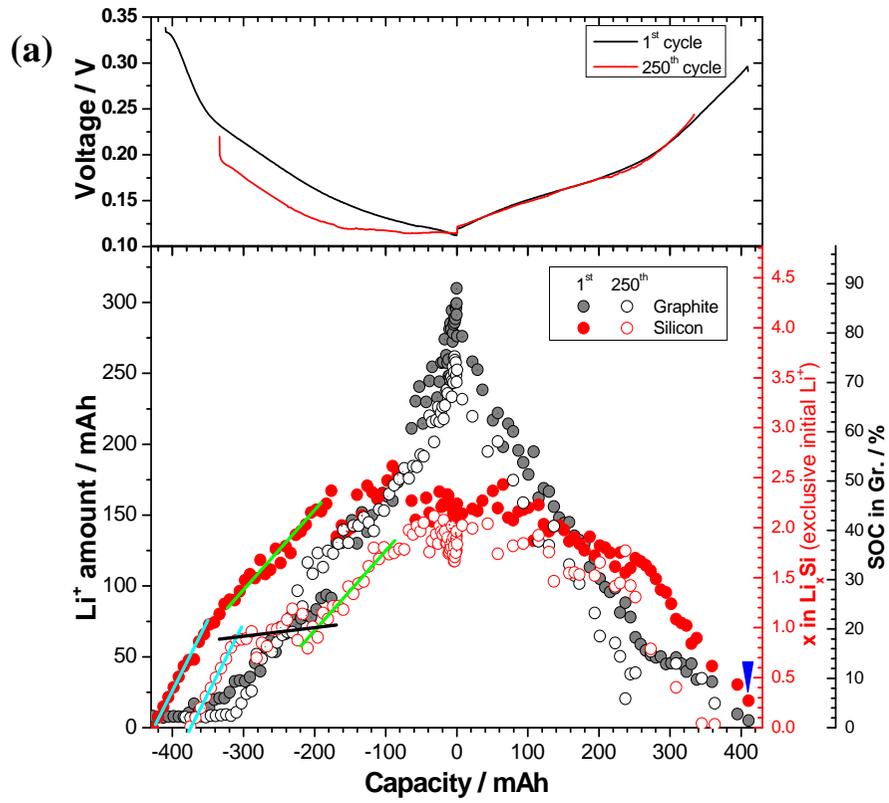
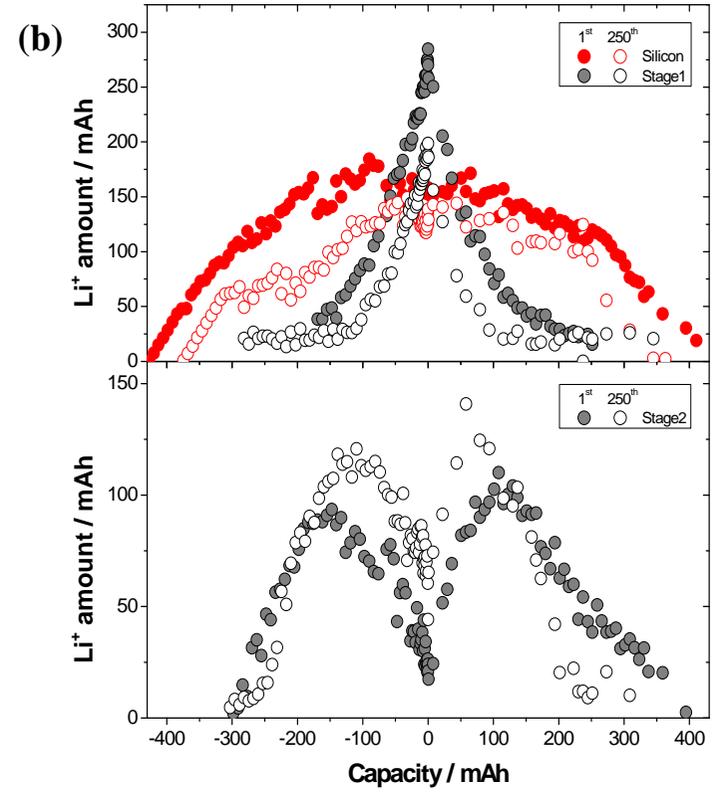

**Figure 2**

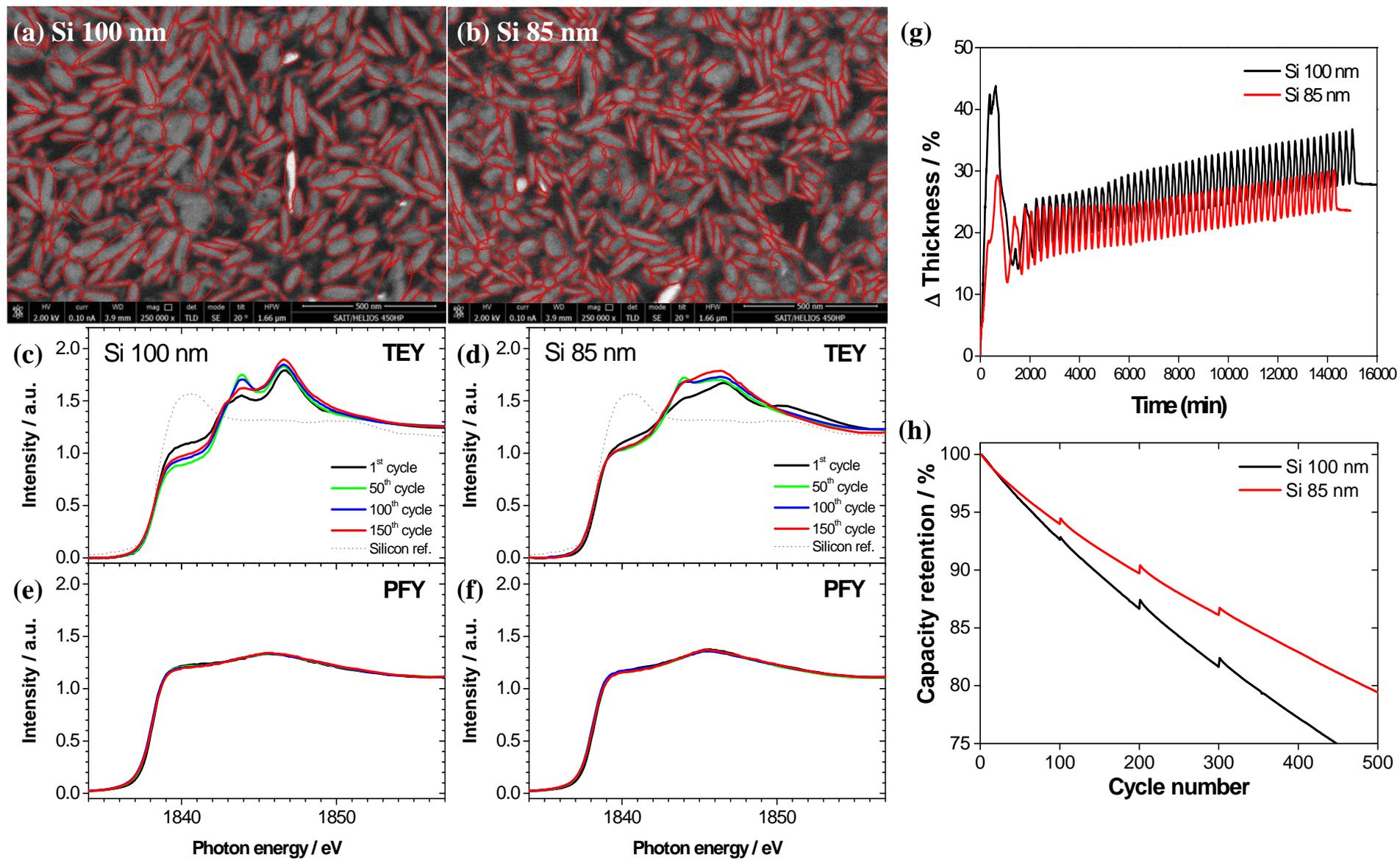

**Figure 3**

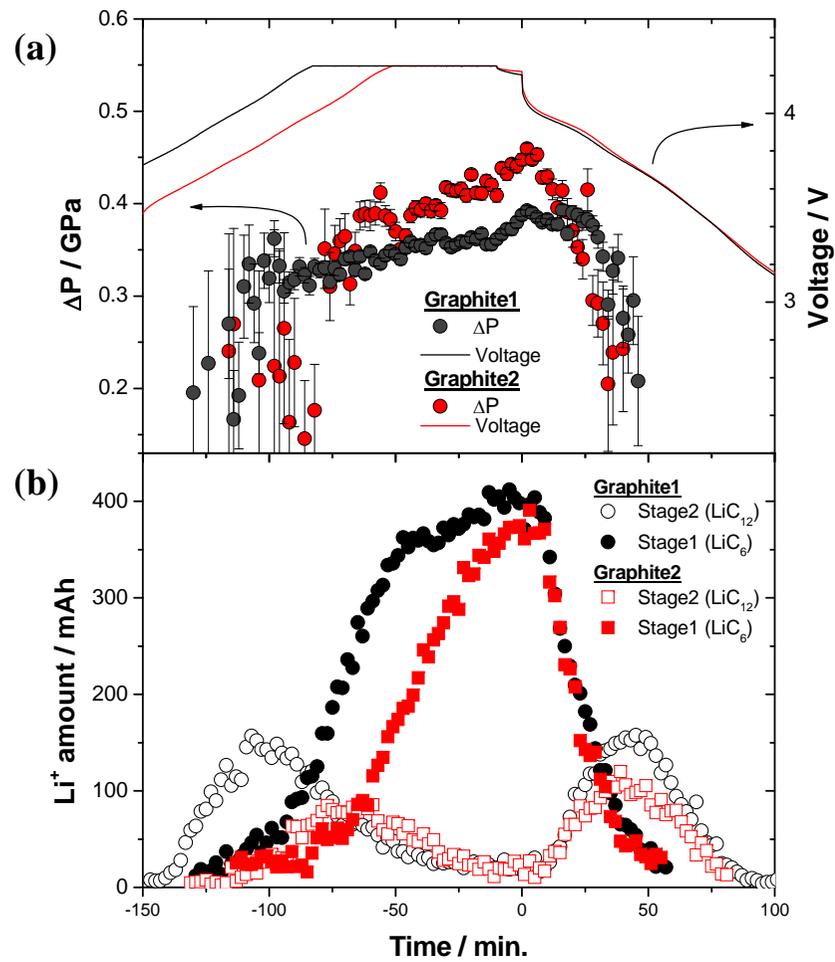
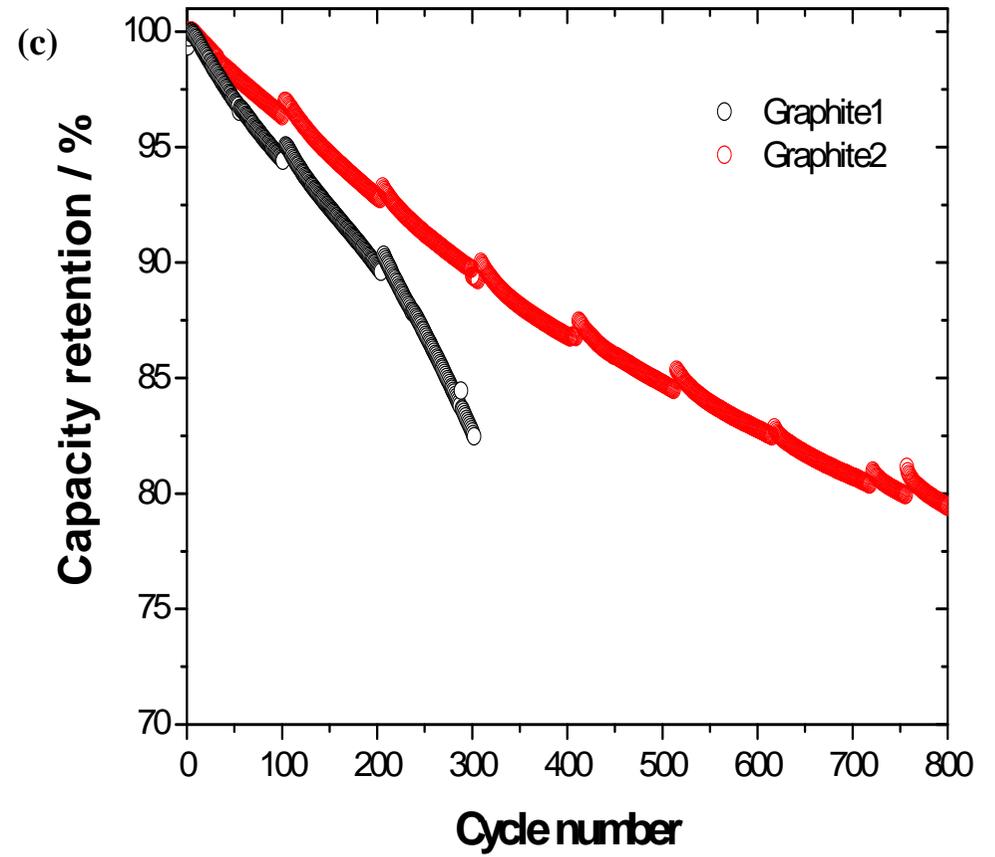

**Figure 4**

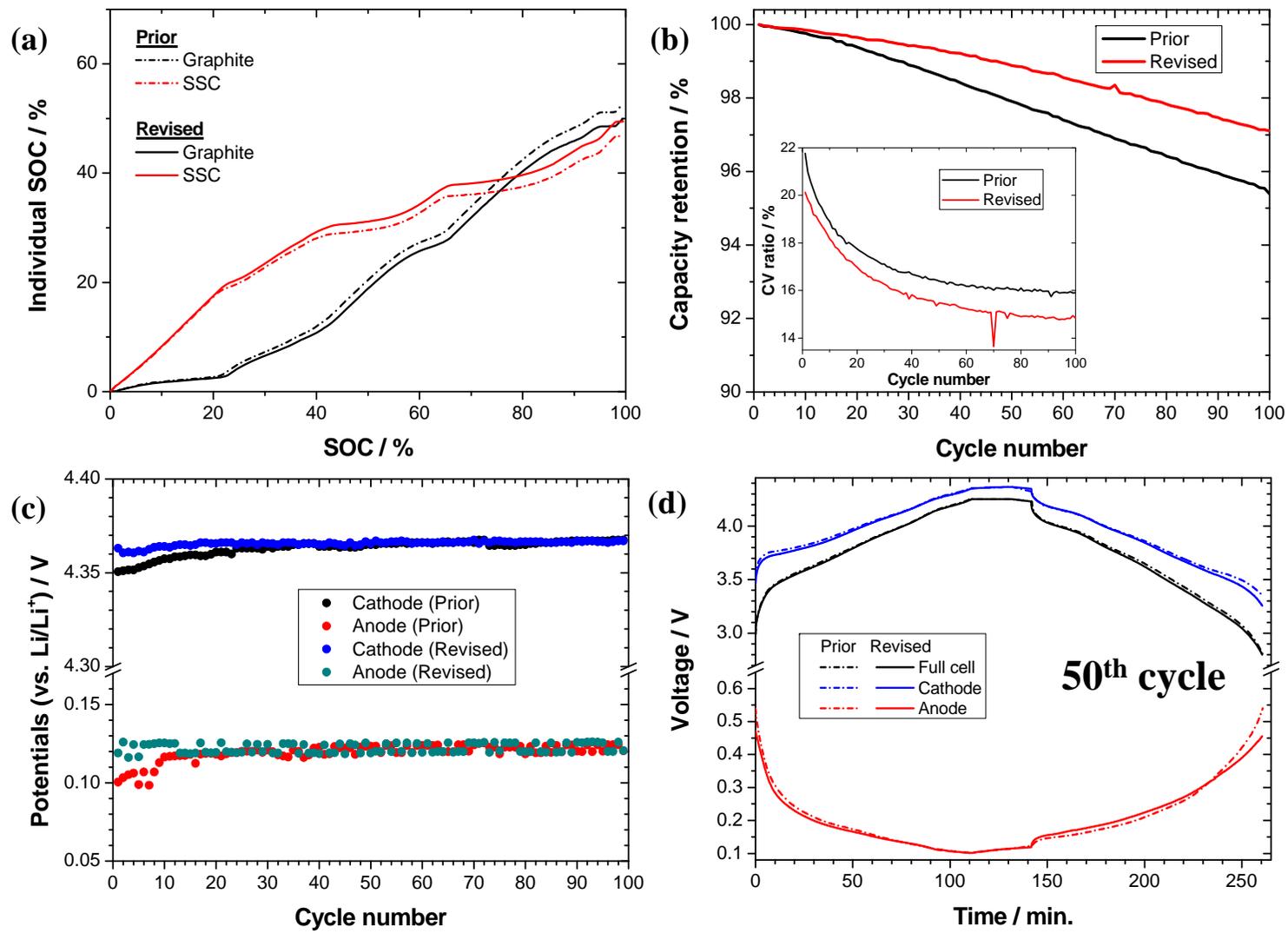

**Figure 5**

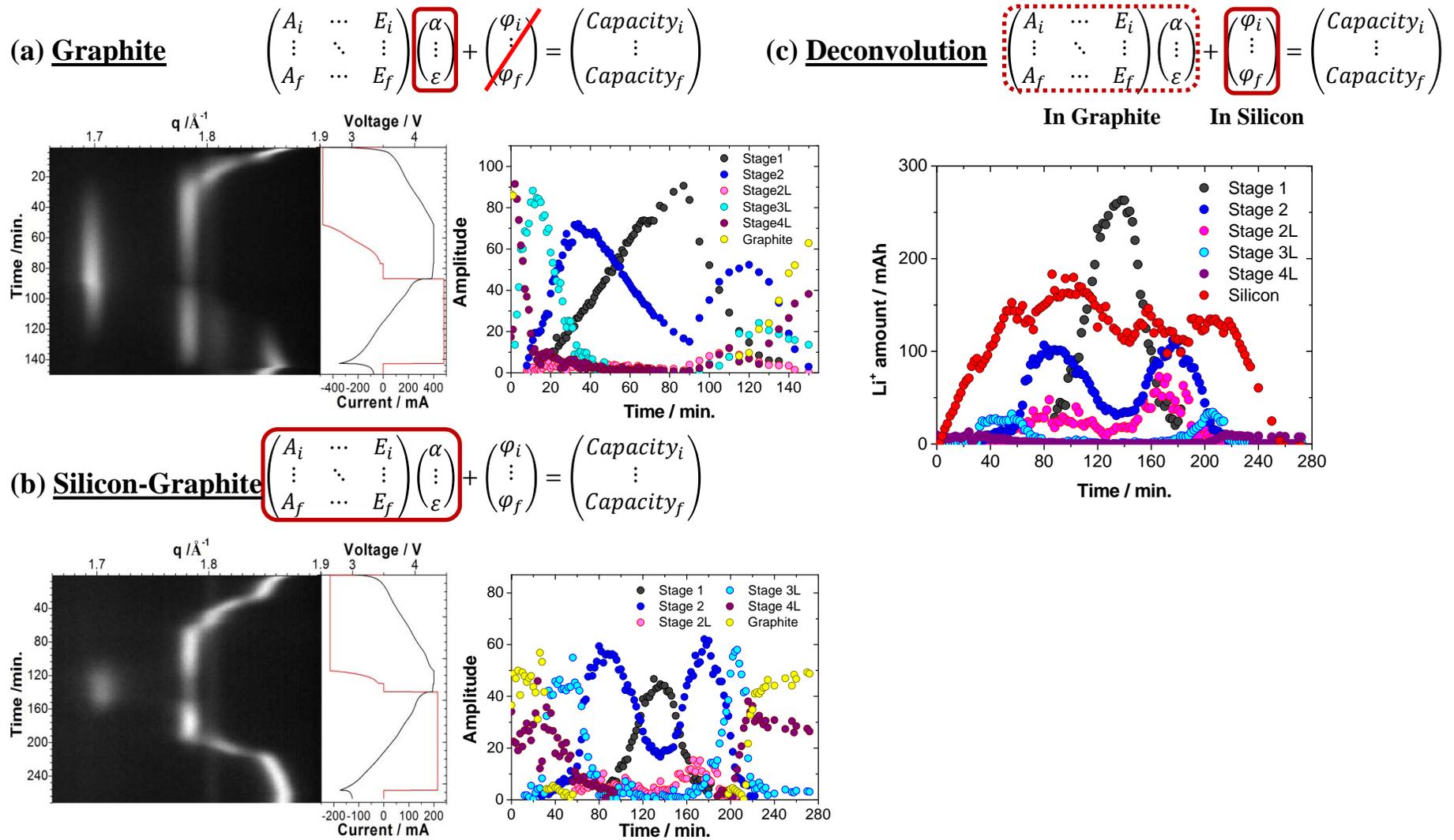

**Figure S1**

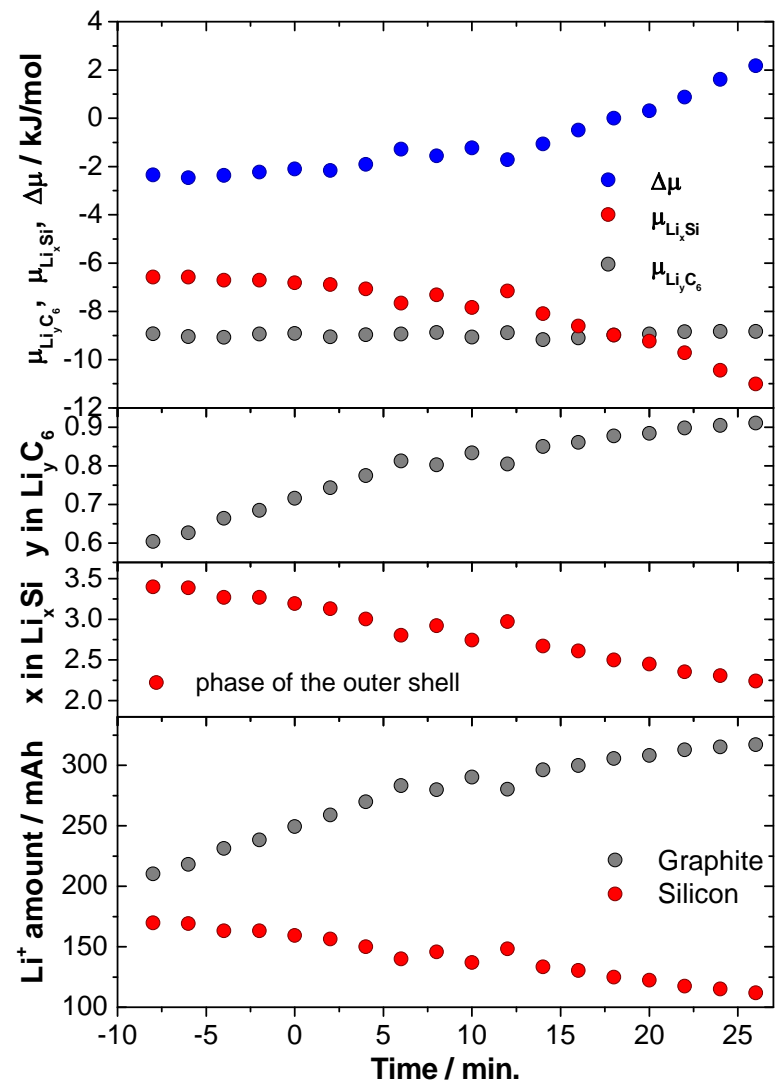

**Figure S2**

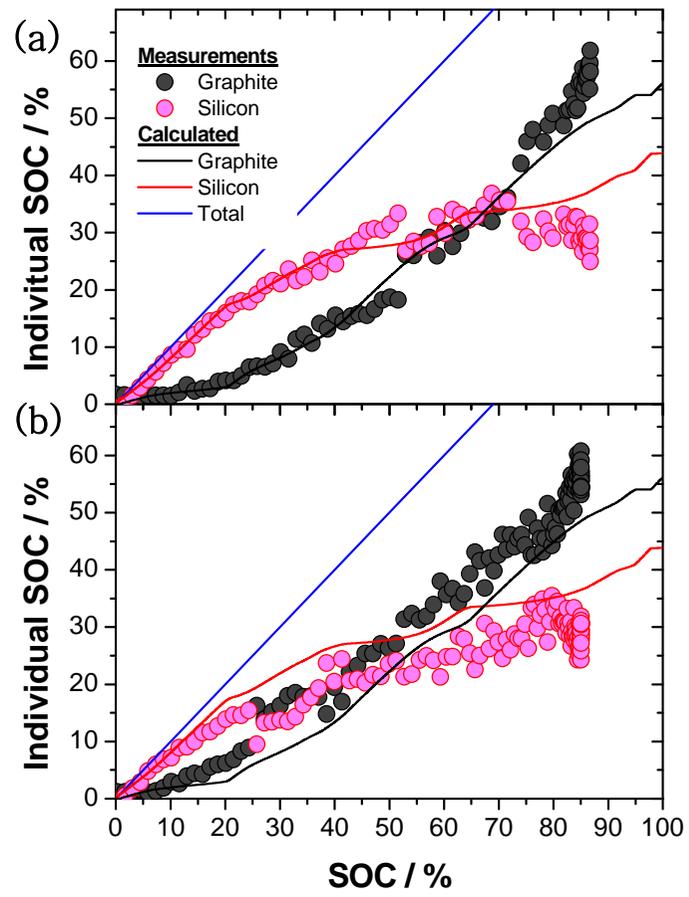

**Figure S3**

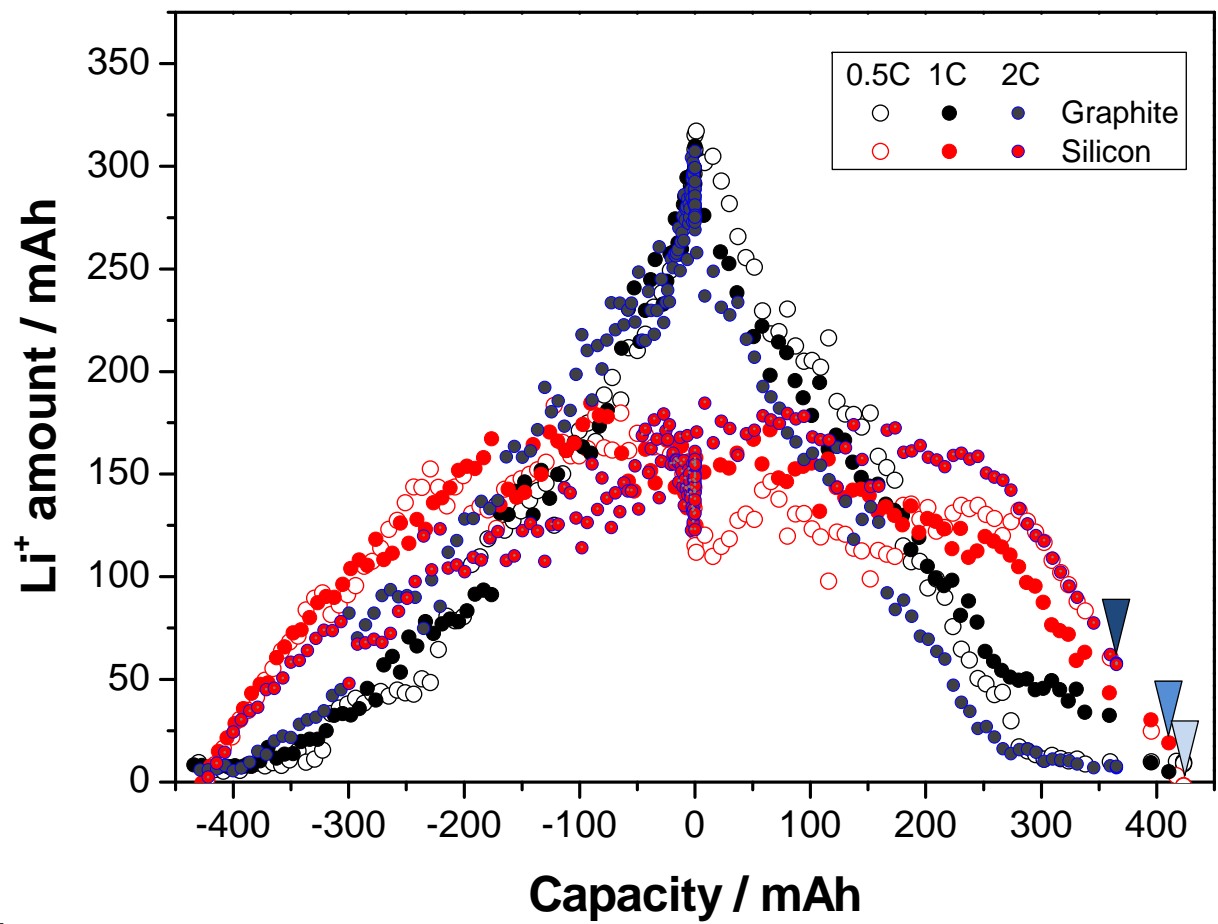

**Figure S4**

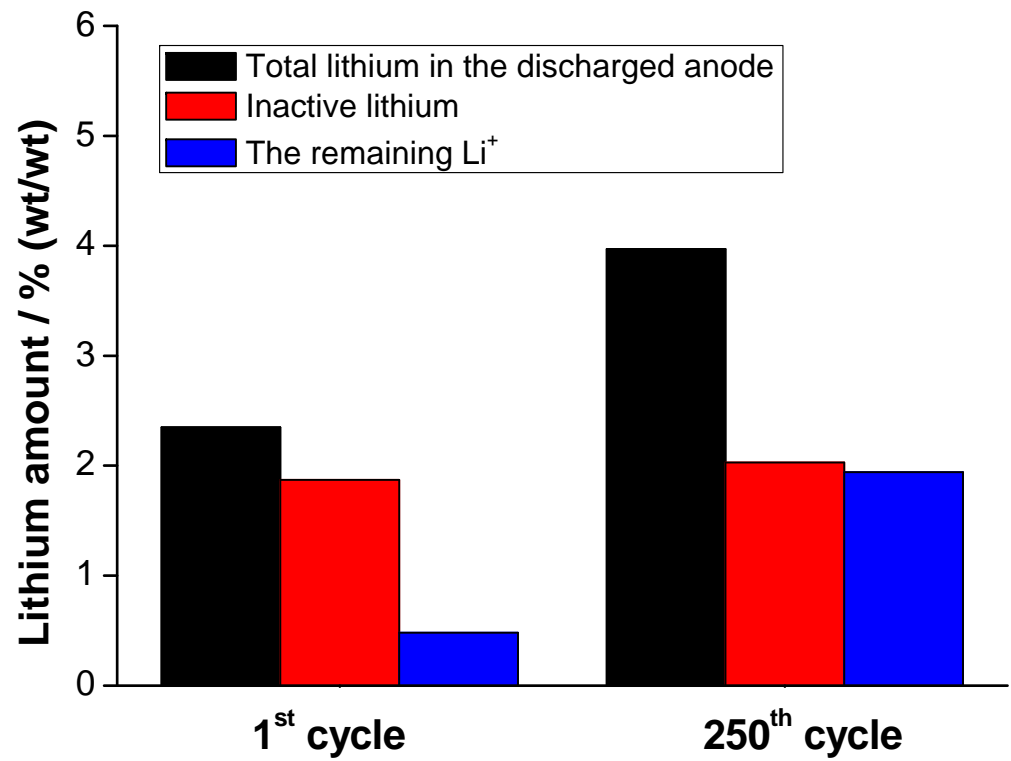

**Figure S5**

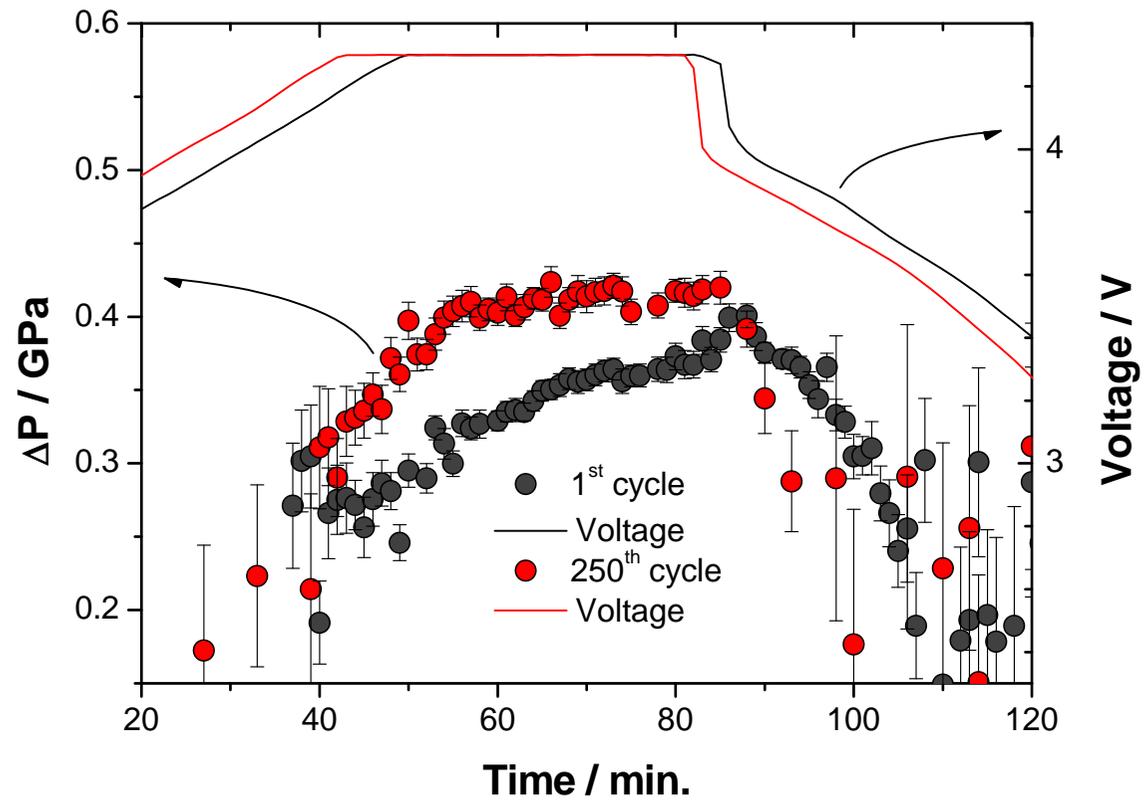

**Figure S6**

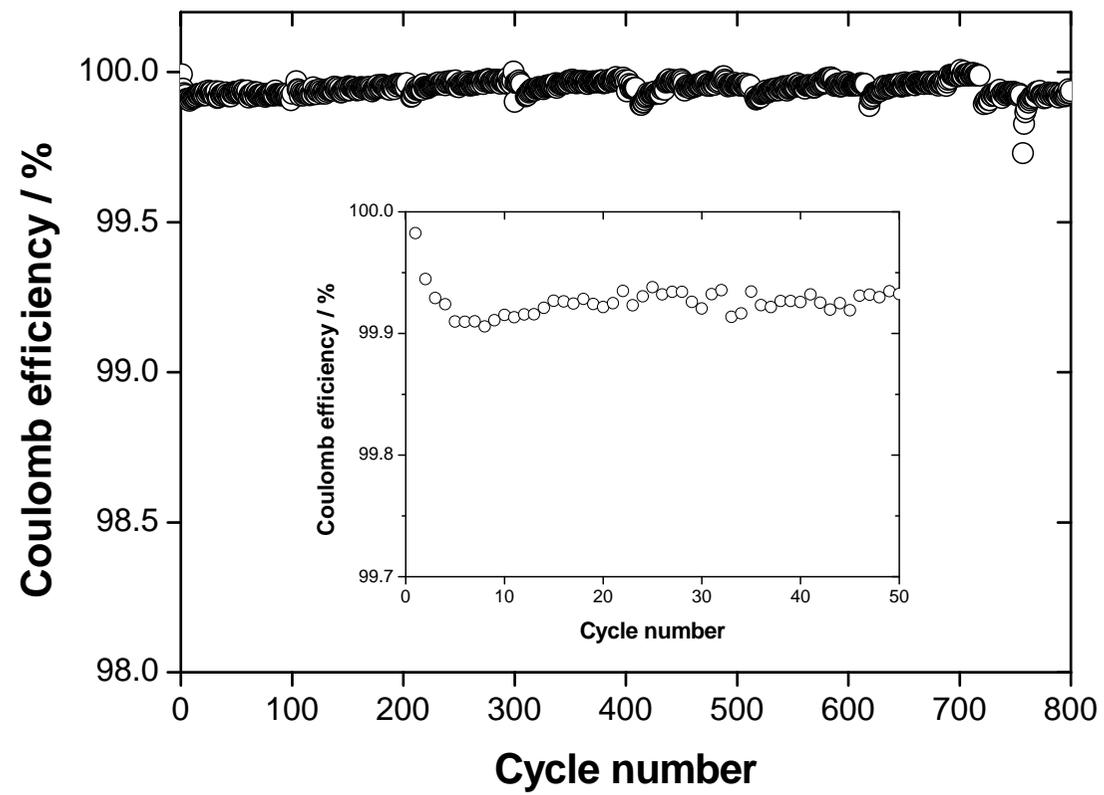

**Figure S7**